\documentclass[10pt,twocolumn,letterpaper]{article}

\usepackage{cvm}
\usepackage{times}
\usepackage{epsfig}
\usepackage{graphicx}
\usepackage{amsmath}
\usepackage{amssymb}
\usepackage{color}
\usepackage{multirow}

\usepackage{booktabs} 
\usepackage{scalefnt}
\usepackage[ruled]{algorithm2e} 

\SetAlFnt{\small}
\SetAlCapFnt{\small}
\SetAlCapNameFnt{\small}
\SetAlCapHSkip{0pt}
\IncMargin{-\parindent}

\DeclareMathOperator*{\argmax}{arg\,max}

\usepackage[pagebackref=true,breaklinks=true,letterpaper=true,colorlinks,bookmarks=false]{hyperref}

\cvmfinalcopy 

\def\cvmPaperID{xxx} 

\ifcvmfinal\fi
\begin{document}

\title{Segmentation-Driven Feature-Preserving Mesh Denoising}


\author{Weijia Wang\footnotemark[1] \\
Deakin University, Australia  
\and
Wei Pan\footnotemark[1] \\
OPT Machine Vision Tech Co., Ltd. (Japan) 
\and
Chaofan Dai \\
OPT Machine Vision Tech Co., Ltd. 
\and
Richard Dazeley \\
Deakin University, Australia 
\and
Lei Wei \\
Deakin University, Australia 
\and
Bernard Rolfe \\
Deakin University, Australia 
\and
Xuequan Lu\footnotemark[2] \\
Deakin University, Australia 
}

\maketitle

\renewcommand{\thefootnote}{\fnsymbol{footnote}} 
\footnotetext[1]{Co-first authors.} 
\footnotetext[2]{Corresponding author.}

\begin{abstract}
    Feature-preserving mesh denoising has received noticeable attention in visual media, with the aim of recovering high-fidelity, clean mesh shapes from the ones that are contaminated by noise. Existing denoising methods often design smaller weights for anisotropic surfaces and larger weights for isotropic surfaces in order to preserve sharp features, such as edges or corners, on the mesh shapes. However, they often disregard the fact that such small weights on anisotropic surfaces still pose negative impacts on the denoising outcomes and detail preservation results on the shapes. In this paper, we propose a novel segmentation-driven mesh denoising method which performs region-wise denoising, and thus avoids the disturbance of anisotropic neighbour faces for better feature preservation results. Also, our backbone can be easily embedded into commonly-used mesh denoising frameworks. Extensive experiments have demonstrated that our method can enhance the denoising results on a wide range of synthetic and real mesh models, both quantitatively and visually.
\end{abstract}

%
%


\section{Introduction}
\label{sec:introduction}

Mesh denoising is a fundamental research problem in geometry processing. The denoised mesh models can be applied to other computer vision tasks such as 3D modelling, computer animation and industrial design. The main challenges lie in the removal of noise while preserving the features (e.g., edges, boundaries and corners) on mesh shapes.

Most existing mesh denoising methods focus on analysing local geometries (e.g., a triangular face and its surrounding neighbours) on triangle mesh surfaces. However, it still remains a challenge in recovering sharp features from noisy mesh models. For example, some methods based on face normals \cite{sun2007fast, Zheng2011, zhang2015guided, Lu2017-efficient} often consider designing small weights for anisotropic surfaces (i.e., where the orientations of neighbouring faces are significantly different) and large weights for isotropic surfaces (i.e., where neighbouring faces have similar normal orientations), in order to preserve sharp features on mesh shapes. Nonetheless, such small weights on anisotropic surfaces still affect the denoising outcomes. There are also anisotropic mesh denoising methods (e.g., \cite{Hildebrandt2004}) attempting to preserve sharp edges and corners on mesh shapes during denoising, but they suffer from recovering features that are contaminated by severe noise. 
A few other methods utilise additional information (e.g., building low-rank matrices based on mesh geometries \cite{lu_lowrank_2020, Li2018}) for mesh denoising. However, such methods are usually very slow due to high computational complexity.

To address the above issues, we introduce a novel segmentation-driven mesh denoising approach to greatly facilitate feature-preserving mesh filtering. 
Our key idea is to partition the mesh surface into segments using an edge-based operator and utilise the segments to guide denoising. Specifically, we cluster triangular faces with similar geometric information (e.g., face normals) into regions and perform region-wise denoising, which assists with eliminating disturbance from anisotropic neighbours and preserving features in the denoised shapes. Furthermore, our segmentation backbone is flexible and can be easily embedded into commonly-used mesh denoising methods. Extensive experimental results have proven that our segmentation approach can greatly boost mesh denoising outcomes in terms of feature preservation, both quantitatively and visually.

Our main contributions are summarised as follows:
\begin{itemize}
\item We design an edge-based segmentation approach to facilitate mesh denoising, which significantly improves the results in regard to feature preservation.
\item Our pipeline's denoising performance is robust on mesh models contaminated by severe noise.
\end{itemize}

\section{Related Work}
\label{sec:relatedwork}

\subsection{Mesh Segmentation}
Mesh segmentation is partitioning the surface of a mesh into meaningful subsets. The methods are usually divided into two broad categories: 1) \textit{semantic} segmentation, which segments the mesh surface into meaningful clusters based on semantic information (e.g., body parts of a human or areas of a city landscape \cite{cgal, grzeczkowicz-2022, Hu_2021_ICCV}); and 2) \textit{geometric} segmentation, which clusters triangular mesh facets based on geometric criteria such as curvatures and normals.

In this paper, we focus on reviewing \textit{geometric} segmentation methods. They can be divided into two classes: \textit{region-based} and \textit{boundary-based} methods. Methods of the former class usually gather regions with similar geometric information (such as curvature or planarity) together, where the most representative methods are K-means and its variants \cite{lian_adaptive_2022}. Inspired by the iterative fitting scheme in K-means, Cohen-Steiner et al. \cite{cohen2004variational} proposed Variational Shape Approximation (VSA), an iterative scheme to reduce distortion error in order to find the best fitting regions to cluster mesh facets. In \cite{achanta2012slic,simari2014fast}, Simple Linear Iterative Clustering (SLIC) technique is adopted to efficiently compute super facets with the K-means approach. Similar to the superpixel concept (i.e., perceptual grouping of pixels) in image processing, triangle faces with similar geometric metrics are grouped into super facets in such works. In addition, there are also other region-based clustering methods such as Mean-shift \cite{comaniciu2002mean}, Medoidshift \cite{sheikh2007mode}, Quick Shift \cite{vedaldi2008quick}, Hierarchical Decomposition \cite{katz2003hierarchical}, Primitive Fitting \cite{attene2006hierarchical} and Random Walks \cite{lai2008fast}. 

Segmentation methods of the latter class detect geometric feature boundaries in input meshes, such that each shape can be divided into different regions based on the boundaries. Relevant methods include Randomised Cuts \cite{golovinskiy2008randomized}, Fuzzy Clustering and Cuts (FCC) \cite{katz2003hierarchical}, Shape Diameter Function (SDF) \cite{shapira2008consistent} and 3D Mesh Scissoring \cite{zheng2011dot,lee2005mesh}. These methods heavily rely on the local geometric information of the input mesh, and may easily fail on complicated or extremely noisy meshes.

In addition to the methods above, recent advances in deep learning lead to data-driven methods for mesh segmentation~\cite{kalogerakis2010learning}. The majority of them deal with the semantic segmentation problem; see~\cite{garcia2017review} for a comprehensive review.

\subsection{Mesh Denoising}
The Laplacian smoothing methods \cite{Vollmer1999, Field1988} are early research works in the field of mesh denoising, which perform isotropic smoothing and are thus fast and efficient. Later, a series of enhanced isotropic methods utilising different techniques such as differential properties, volume preservation and pass frequency-controlling were introduced \cite{liu2002,kim2005,Nehab2005,Nealen2006,su2009} to improve mesh denoising outcomes. Nevertheless, while such methods are effective for noise removal, their isotropic nature may blur or shrink sharp features on the mesh shapes.

In order to alleviate such feature-blurring problems, anisotropic methods started to emerge. An early anisotropic denoising work by Hildebrandt and Polthier \cite{Hildebrandt2004} utilises mean curvature flow to denoise mesh shapes. To better preserve features, two-step methods such as bilateral filtering techniques \cite{Lee2005, Zheng2011} and others \cite{sun2007fast,zhang2015guided,Lu2017-efficient, lu_lowrank_2020, chen_uniform_point_distri_2022} were proposed in the following years. Such methods involve two steps: normal smoothing and vertex updating, which have demonstrated promising outcomes for robust, feature-preserving mesh denoising. In recent years, some researchers also attempt to classify vertices and faces in order to distinguish features during mesh denoising \cite{Wei2015,Fan2010,Bian2011,Wang2012,Zhu2013,Wang2009,Wei2017}. Nonetheless, such classification strategies mainly focus on local neighborhoods and are usually sensitive to noise. To mitigate the issue, Lu et al. \cite{Lu2016,Lu2017-L1median} proposed a pre-filtering technique before denoising as a remedy solution, such that the impact brought by excessive noise is significantly reduced.

Another stream of anisotropic mesh denoising method focuses on the sparse perspective, as feature vertices can be computed by solving linear sparse systems. For example, He and Schaefer \cite{He2013} proposed an $L_{0}$-minimisation framework, which utilises a discrete differential operator to preserve mesh features during denoising. While this method is straightforward, the minimisation process (as solving a sparse system) is non-convex and slow. To improve this, Zhao et al. \cite{Zhao2018} introduced an alternating optimisation strategy to perform $L_0$-minimisation, which consists of 2 steps (i.e., updating vertex positions and face normals). With similar inspiration and backbones, Lu et al. \cite{Lu2017-L1median} introduced an $L_1$-minimisation method to preserve mesh features during denoising. Recently, by constructing half window of the local neighborhood for each vertex, Pan et al. \cite{pan2020hlo} proposed a half-kernel Laplacian operator to reduce the damages on features while removing noise. However, while this method is fast and effective, it has limited capability for sharp edge preservation on CAD-like models (i.e., models that have sharp edges separating smooth regions, such as cubes and octahedrons).

Over the years, there are works attempting to utilise mesh segmentation results to guide mesh denoising. For instance, Vieira and Shimanda \cite{vieira_surface_2005} proposed a region-growing algorithm to segment noisy mesh surfaces and then smooth them out. While this is capable on scanned meshes, it can hardly preserve sharp features among the regions. Later, Huang and Ascher \cite{huang_surface_2008} proposed a vertex classification technique to segment meshes and guide denoising. However, it requires manually tuning the number of clusters based on the input shapes and does not achieve good feature-preserving results on scanned models. Lagrand et al. \cite{Legrand:2019filtered} proposed a variant of Quick Shift to automatically optimise the number and distribution of mesh segments, but the clustering result is obtained based on filtered meshes. Recently, Wang et al. \cite{wang_mumford-shah_2022} utilised Mumford-Shah function to obtain outlines for each noisy mesh and perform feature-preserving denoising. Nevertheless, the segmentation and denoising outcomes may still be unsatisfactory at high noise levels. With the benefits of mesh segmentation in mind, a method that can utilise the advantages of segmentation and guide feature-preserving denoising is thus greatly desired.

\section{Method}
\label{sec:method}

\subsection{Method Overview}

\begin{figure}[t]
	\centering{
		\includegraphics[width=\columnwidth]{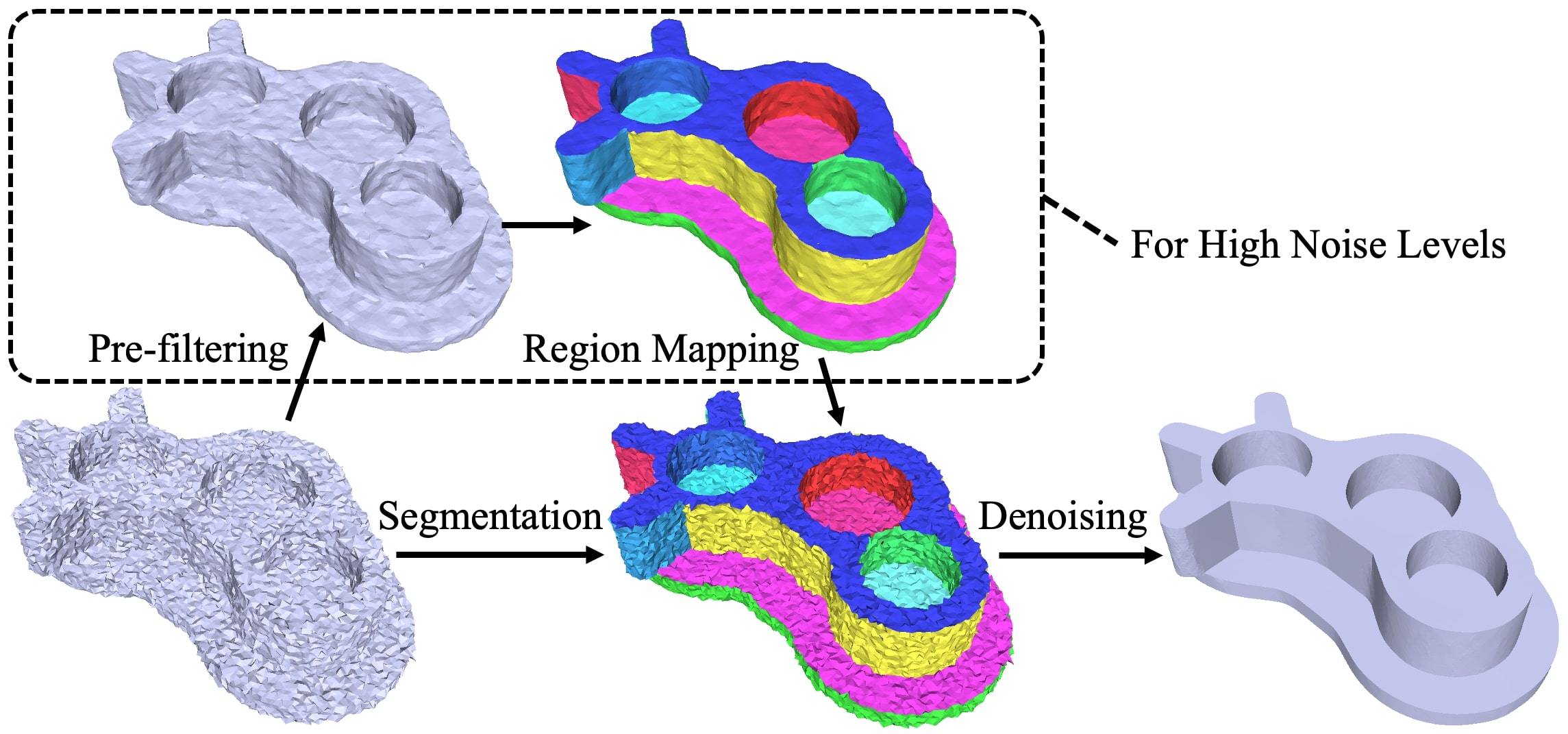}
	}
	\caption{Our mesh denoising pipeline.} 
	\label{fig:pipeline}
\end{figure}

Fig.~\ref{fig:pipeline} provides an overview of our segmentation-driven mesh denoising method. The key idea is segmenting the triangular faces on a noisy mesh into regions based on the mesh's features (e.g., edges and corners). Each region is then denoised separately without being affected by neighboring facets in other regions. In this way, the features can be more effectively preserved in the denoised mesh. Note that for meshes corrupted with relatively higher levels of noise, we first employ an additional pre-filtering step to remove excessive noise. We then perform region segmentation on the pre-filtered mesh so that more accurate segmentation results can be obtained. Finally, we map the segmented regions back to the original unprocessed noisy mesh and utilise the regions to guide mesh filtering. We elaborate the pre-filtering step in Sec.~\ref{sec:pre-process}, demonstrate our segmentation approach in Sec.~\ref{sec:segmentation}, and introduce the denoising techniques in Sec.~\ref{sec:method-denoising}.

\subsection{Pre-filtering}  
\label{sec:pre-process}
We aim to obtain high-quality segmentation results to guide our subsequent denoising operations. For meshes with low noise, we directly perform segmentation on them as their triangular faces are not severely distorted. Nevertheless, the segmentation task becomes extremely difficult on meshes corrupted with high noise levels (i.e., 30\% of the mesh's average edge length as per our experimental results), where the face orientations are severely distorted or even flipped. It is extremely difficult to directly perform area segmentation on such mesh shapes, as the severely distorted triangular faces do not provide indicative information for our segmentation process. Thus, we introduce this \textit{pre-filtering} step to remove excessive noise to obtain better segmentation results. Inspired by \cite{Lu2016}, we formulate this pre-filtering step as a convex optimisation problem, where we estimate new vertex positions for the mesh. The objective function is formulated as
\begin{equation}\label{eq:prefiltering}
\resizebox{\hsize}{!}{
    $ \min \sum_i ||\tilde{p}_i-p_i||_2^2 + \alpha\sum_e w(e)||D(e)||_2^2 + \beta\sum_e w(e)||R(e)||_2^2 $,
    }
\end{equation}
where $p_i$ is the original $i$-th vertex of the input mesh and $\tilde{p}_i$ is its unknown denoised position. $D(e)$ and $R(e)$ are the area-based edge operator and the regulariser that were originally defined in \cite{He2013}. Both $D(e)$ and $R(e)$ are weighted by a Gaussian function $w(e) = e^{-\left ( \frac{\theta }{\sigma _{\theta }} \right )^{2}}$; the function itself is designed to preserve features during the pre-filtering step, where $\theta$ represents the angle formed by each two adjacent faces' normals, and $\sigma_\theta$ is a scaling threshold for normal similarity. In addition, $D(e)$ and $R(e)$ are respectively multiplied by two global weighting parameters, $\alpha$ and $\beta$. 

We solve Eq.~\ref{eq:prefiltering} as a sparse linear system. In specific, we firstly conduct 1 unweighted iteration (i.e., without the term $w(e)$), followed by 2 weighted iterations (i.e., with $w(e)$) as per \cite{Lu2016}. Also, based on our experimental results, we set $\alpha$, $\beta$ and $\sigma_\theta$ from $w(e)$ in Eq.~\ref{eq:prefiltering} to 0.2, 0.1 and 30 degrees respectively, as such values lead to satisfactory segmentation results on our test shapes.


\subsection{Edge-based Region Growing Segmentation}
\label{sec:segmentation}
\textbf{Segmentation.} We propose to use edge metrics to perform mesh segmentation. An intuitive way is to utilise the face normal information on the mesh surface to distinguish features. To do so, we set a global threshold $N_{thr}$ which is defined as 
\begin{equation}
\label{eq:global-diff}
    N_{thr} = \frac{\sum_{e\in E} \cos(n_i, n_j)}{\left| E \right|},
\end{equation}
where $\left| E \right|$ is the total number of edges in the input mesh, $n_i$ and $n_j$ are the normals of the two faces sharing edge $e$, and $\cos(n_i, n_j)$ is the cosine value of the angle formed by $n_i$ and $n_j$.

\begin{figure} [t]
\centering
\includegraphics[width=\columnwidth]{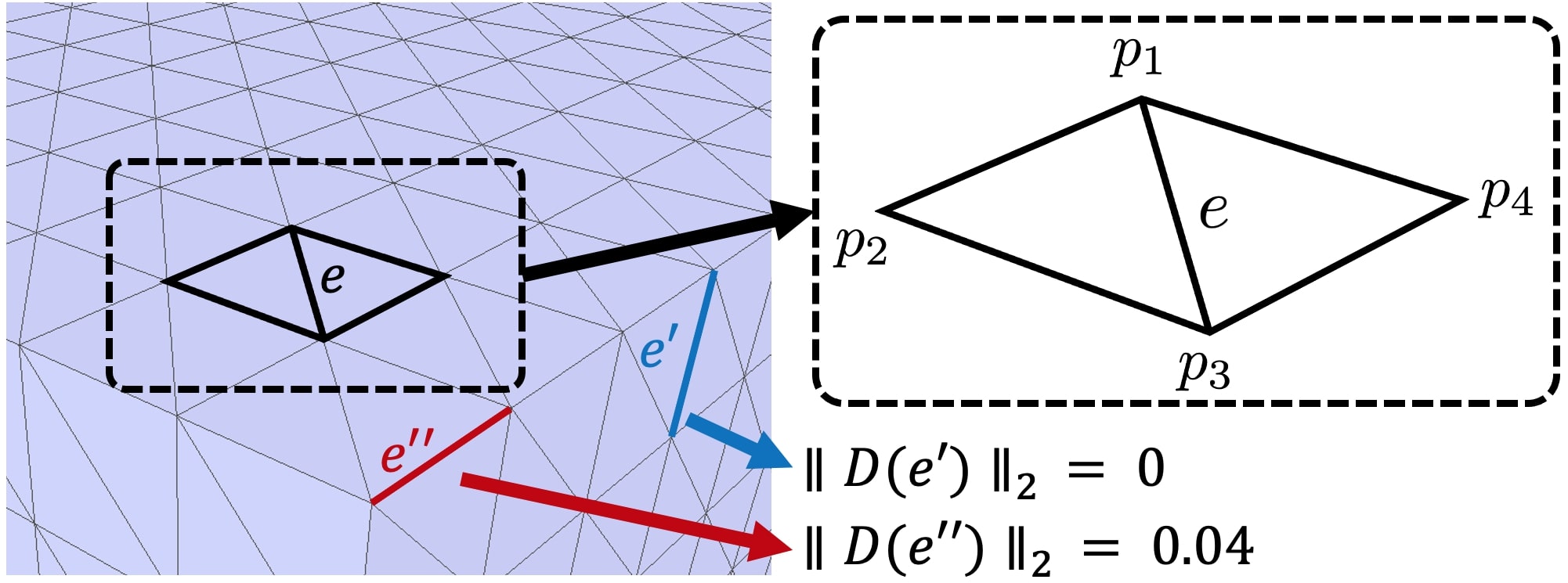}
\caption{Demonstration of the edges and vertices. For any edge $e$, we denote the order of the 4 adjacent vertices as per \cite{He2013}. Also, for edge $e'$ on a flat surface and edge $e''$ on a sharp feature, the $L_2$-norms of their edge operators are 0 and 0.04 respectively in this example.}
\label{fig:vertex_order}
\end{figure}

Nevertheless, since noise distorts the faces' orientations, this global threshold may give false features (as shown in Sec.~\ref{subsec:segmentation-methods}). To alleviate this issue, we introduce $D(e)$ from Eq.~\ref{eq:prefiltering} as an edge operator to distinguish features locally. As per the definition in \cite{He2013}, for any two adjacent triangles sharing an edge $e$, we assume the surrounding four vertices are ordered as per Fig.~\ref{fig:vertex_order}. Based on this order, $D(e)$ is defined as

\begin{equation}
\label{e1}
\begin{aligned}
\resizebox{0.99\hsize}{!}{
$D(e) = 
\left[
\begin{array}{c}
\frac{\triangle_{123}(p_4-p_3)\cdot(p_3-p_1)+\triangle_{134}(p_1-p_3)\cdot(p_3-p_2)}{(\begin{vmatrix}p_3-p_1\end{vmatrix})^2(\triangle_{123}+\triangle_{134})}\\ 
\frac{\triangle_{134}}{\triangle_{123}+\triangle_{134}} \\ 
\frac{\triangle_{123}(p_3-p_1)\cdot(p_1-p_4)+\triangle_{134}(p_2-p_1)\cdot(p_1-p_3)}{(\begin{vmatrix}p_3-p_1\end{vmatrix})^2(\triangle_{123}+\triangle_{134})} \\
\frac{\triangle_{123}}{\triangle_{123}+\triangle_{134}} \\ 
\end{array}
\right]^\mathrm{T}
\left[
\begin{array}{ccc}
p_1\\
p_2\\
p_3\\
p_4\\
\end{array}
\right]$,
}
\end{aligned}
\end{equation}
where $\triangle_{123}$ denotes the area of the triangle formed by $p_1$, $p_2$ and $p_3$, and $\triangle_{134}$ is the area of the triangle formed by $p_1$, $p_3$ and $p_4$. 

$D(e)$ itself is a $1 \times 3$ vector, which essentially describes the feature or non-feature property of a specific edge $e$. We calculate its $L_2$-norm to determine whether or not $e$ is a feature edge: the value should be $0$ when the adjacent triangle pair forms a flat dihedral angle, and increases if the dihedral angle becomes smaller (as shown in Fig.~\ref{fig:vertex_order}). Based on the properties of $D(e)$, we introduce a global threshold $D_{thr}$ to decide feature edges, where an edge is a feature edge if the $L_2$-norm of its $D(e)$ is greater than $D_{thr}$ (i.e., $||D(e)||_2 > D_{thr}$). Unfortunately, merely relying on the term $D(e)$ is also very fragile for detecting feature edges, as shown in Sec.~\ref{subsec:segmentation-methods}. Thus, for our segmentation technique, we use $N_{thr}$ in Eq.~\ref{eq:global-diff} as a global prior term and use $D(e)$ as a complementary term. These two terms together form our edge detection metric for our region-growing segmentation algorithm.



\begin{figure} [t]
	\begin{center}
			\includegraphics[width=\columnwidth]{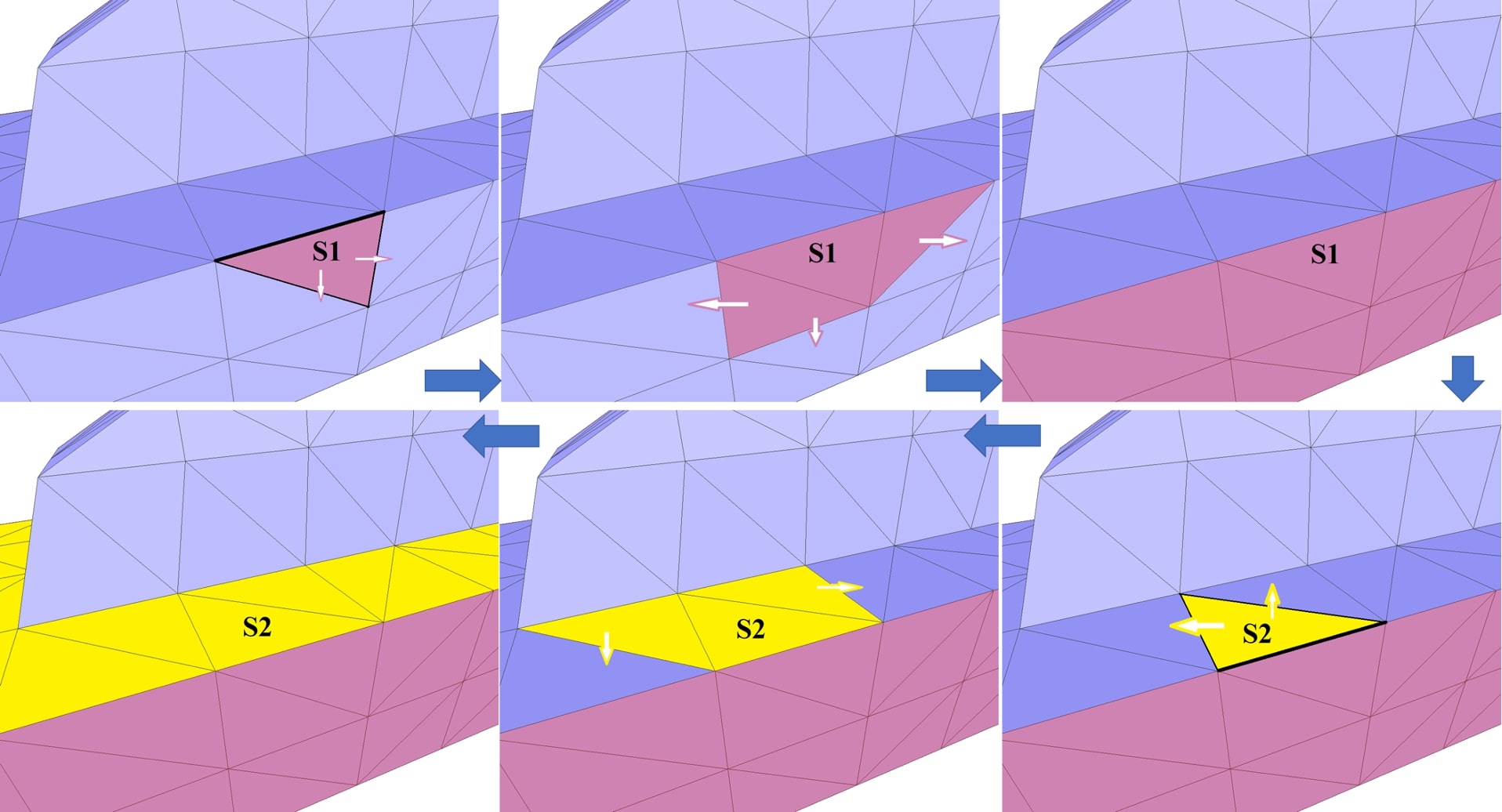}
	\end{center}
	\caption[front] 
	{ \label{fig:region-grow} 
		Demonstration of our edge-based region growing segmentation method. The seeding face $S1$ will be expanded based on edge-connected normal angles and the edge operators. When the expansion of $S1$ stops, we randomly select an unprocessed face $S2$ and repeat the expansion process, until all faces on the mesh are clustered.}
\end{figure}

Fig.~\ref{fig:region-grow} shows our region segmentation process. We first set a seeding triangle face in the input mesh and assign a label to it. For each edge of the seeding face, we compute the cosine of the normal angle formed with the corresponding edge-connected face (i.e., $\cos(n_i, n_j)$), as well as the edge's $D(e)$. We then determine if that edge-connected face belongs to the same cluster as the seeding face, by comparing $\cos(n_i, n_j)$ with $N_{thr}$ and $||D(e)||_2$ with $D_{thr}$. If $\cos(n_i, n_j)$ is greater than $N_{thr}$ or $||D(e)||_2$ is less than $D_{thr}$, we assign the cluster label of the seeding face to that connected triangle face. The newly-clustered faces with the same label are all regarded as seeding faces. We further calculate $\cos(n_i, n_j)$ and $||D(e)||_2$ for the remaining edges of the newly-clustered faces, and repeat this procedure to keep expanding the seeding cluster until no more satisfactory adjacent triangles can be found. When this happens, we randomly select a new seeding face from the remaining unclustered faces of the input mesh and execute the above procedure again. Our segmentation procedure finishes when all faces on the mesh are clustered.

\SetKwRepeat{Do}{do}{while}
\begin{algorithm}[t]
\SetKwInput{KwData}{Input}
\SetKwInput{KwResult}{Output}
  \KwData{Mesh with low noise or pre-processed mesh, and the threshold $D_{thr}$\; }
  \KwResult{Mesh partitioned in segments\;}
  Compute $N_{thr}$\;
    \Repeat{all faces are clustered}
    {
      Randomly select an unprocessed face $F_{i}$ as a seed for a new cluster $C$\;
      Get $F_{i}$'s edge-connected neighbors $\{F_{j}\}$ and corresponding edges $\{e_{j}\}$\;
  	      \While{$\cos(n_i, n_j)>N_{thr}$ \textbf{or} $||D(e_{j})||_2<D_{thr}$}
  	      {
  	      cluster $F_{j}$ into $C$\;
 	      mark $F_{j}$ as a new seed\;
      }
    }
  Refine the segments\;
  \caption{Region Growing Segmentation}
  \label{alg:region-grow-algo} 
\end{algorithm}



\begin{figure} [t]
\centering
			\includegraphics[width=0.8\columnwidth]{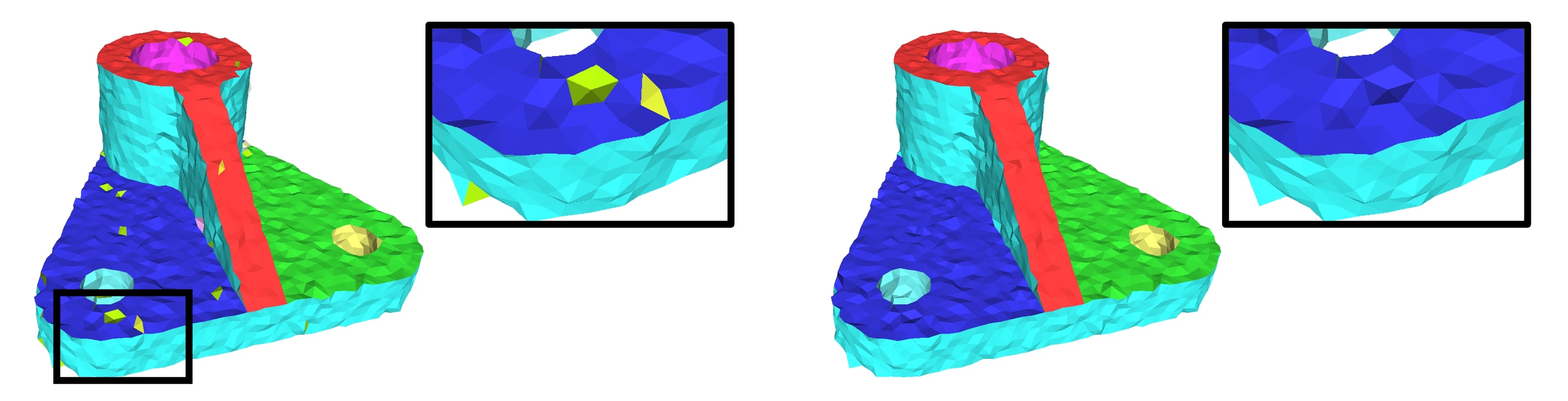}\\
			\makebox[0.4\columnwidth]{(a) }
			\makebox[0.4\columnwidth]{(b) }\\
\caption{Segmentation results (a) before and (b) after region refinement. }
\label{fig: refine-demo}
\end{figure}

\textbf{Region Refinement.} Small clusters are sometimes observed after the region-growing segmentation process, as shown in Fig.~\ref{fig: refine-demo}(a). Such small areas are leftover triangular faces which were not clustered into their surrounding regions due to noise. We further find that such small clusters usually pose negative impacts to mesh denoising, as discussed later in Sec.~\ref{subsec:refinement-compare}. Thus, we identify such overly-small clusters and merge them into nearby ones. In our experiments, we empirically set this threshold to 50 faces, as it can identify the undesired small clusters being left due to noise. Specifically, for each triangle face in the small cluster, we calculate the cosine value of the current face normal and each of its 2-ring neighborhood face normal, and sum the cosine within the same cluster. For simplicity, the cluster label which produces the greatest sum is assigned to this triangle face. The function is defined as
\begin{equation}
    \argmax_k \sum_{j\in S(i)} \cos(n_i,n_j),
\end{equation}
where $k$ indicates the desired cluster label, $n_i$ is the current face normal, and $j \in S(i)$ represents any face with index $j$ in the 2-ring neighborhood face set $S(i)$, with $n_j$ as its normal. The desired label $k$ is chosen from all candidate faces in the face set $S(i)$. The segmentation result after the region refinement process is demonstrated in Fig.~\ref{fig: refine-demo}(b), where the small clusters are fused into surrounding regions.


\begin{figure} [t]
	\begin{center}
			\includegraphics[width=0.8\columnwidth]{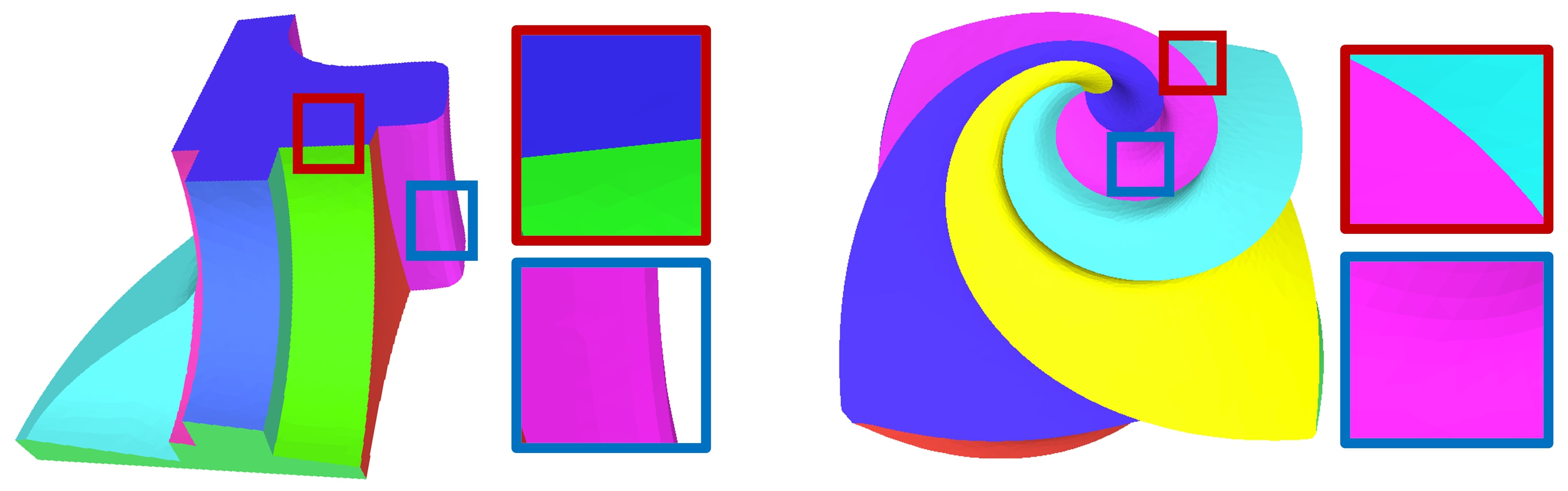}
	\end{center}
	\caption{Segmentation on shapes with mixed types of surfaces. Sharp edges (in red frames) are distinguished by our algorithm using two colours, while smooth regions (in blue frames) are marked in the same colour. }
	\label{fig:segment-complex} 
\end{figure}

The full pipeline of our edge-based region growing segmentation algorithm is summarised in Alg.~\ref{alg:region-grow-algo}. Overall, the term $N_{thr}$ provides a global threshold for guiding feature identification during region segmentation. Based on this global prior term, the use of $D_{thr}$ helps with distinguishing mixed types of surfaces, such as shapes with combinations of sharp edges and soft curves. Fig.~\ref{fig:segment-complex} shows the segmentation performance on two mesh shapes, Fandisk and Octaflower, where we show close-ups of sharp regions in red frames and smooth regions in blue frames. The segmentation threshold $D_{thr}$ is set to 0.03 and 0.0001 respectively for each shape. As can be seen, smooth surfaces are clustered into the same region while sharp boundaries are distinguished by two colours. 

\subsection{Mesh Denoising}
\label{sec:method-denoising}

\begin{figure} [t]
\centering
\includegraphics[width=0.7\columnwidth]{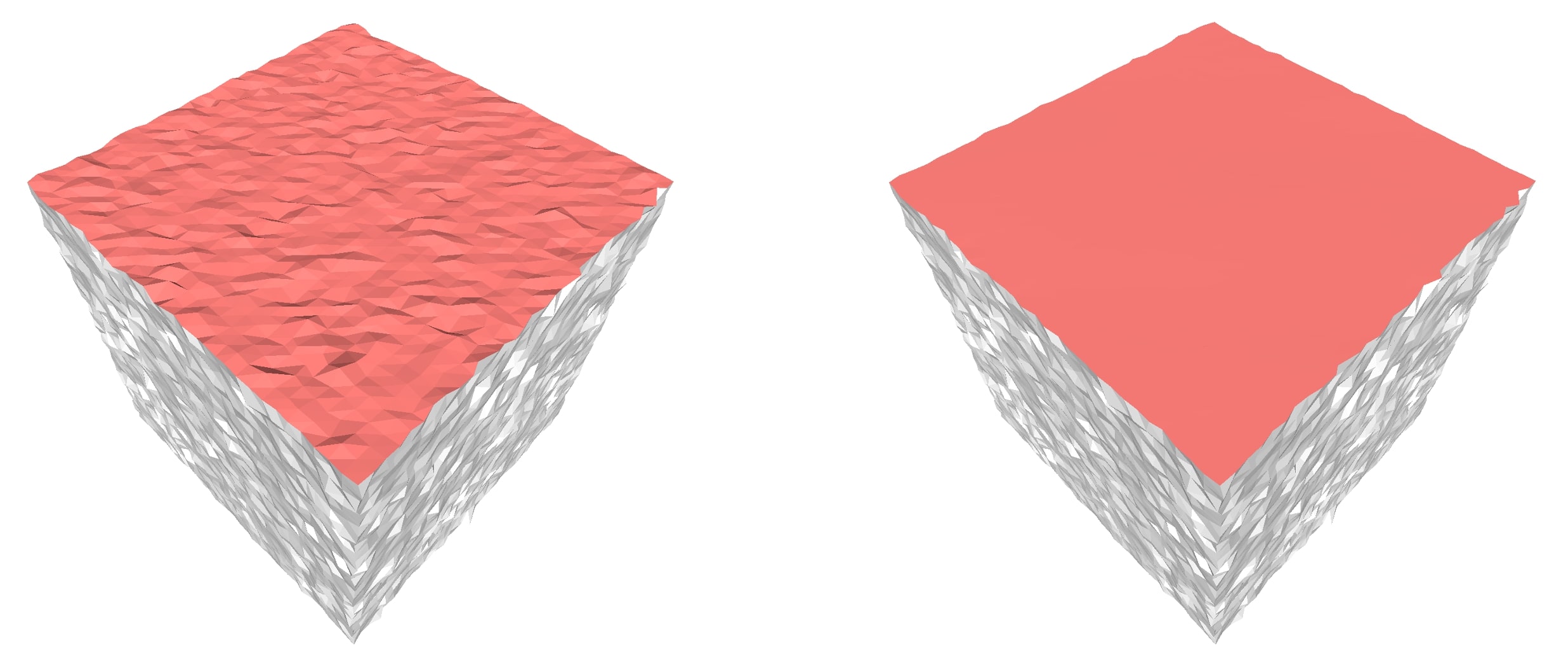}
\caption{Demonstration of region-based denoising, where faces in each region (marked in red) in the noisy shape (left) are denoised (right) without considering any neighbours outside the region.}
\label{fig:region-denoise}
\end{figure}

Our segmentation algorithm partitions the triangular faces on the mesh shape into regions, where each region excludes anisotropic neighbouring faces. We then constrain the neighbouring faces and update their normals and vertices within each region. This step excludes negative influence from the neighbouring faces on anisotropic surfaces that may affect the denoising results and prevent feature preservation. The procedure is demonstrated in Fig.~\ref{fig:region-denoise}, where all faces within a specific region (marked in red) are denoised without including any adjacent faces outside the region. In this respect, many local-based mesh denoising techniques can gain benefits from our segmentation step, such as \cite{sun2007fast, Zheng2011, zhang2015guided, Lu2017-L1median}. Experimental results demonstrate that our method can reach state-of-the-art results, which will be elaborated in the upcoming section.

\section{Experimental Results}
\label{sec:results}

We select the following methods for comparison: $L_0$-minimisation (L0) \cite{He2013}, the Bilateral Normal Filter (BNF) \cite{Lee2005}, the Unilateral Normal Filter (UNF) \cite{sun2007fast}, the Guided Normal Filter (GNF) \cite{zhang2015guided}, the $L_1$-median Filter (L1) \cite{Lu2017-L1median}, the Half-kernel Laplacian Operator (HLO) \cite{pan2020hlo} and Total Generalized Variation (TGV) mesh denoising \cite{liu_tgv_2021}. We select them since they are the state-of-the-art works over the years, and their source code or executables are available online. For our denoising step, we embed four open-source methods into our segmentation backbone: BNF, UNF, GNF and L1. 

Similar to our pipeline, some of the comparison methods also utilise the anisotropic denoising idea. 
For instance, the bilateral filter in \cite{Lee2005} utilises anisotropic diffusion during the denoising task. Similarly, the guidance normal construction process in \cite{zhang2015guided} is inspired by anisotropic denoising (i.e., judging a face patch's normal consistency). Nevertheless, experimental results show that such methods still have limitations on feature preservation and our segmentation-driven backbone can achieve more desirable effects on the denoised meshes. 

\subsection{Parameter Setting}
The parameters of the methods being used for our comparison experiments are summarised in Table~\ref{table:tablepara}. For all methods, we use the recommended parameters and carefully tune them to obtain desired outputs. Compared with other methods, ours only has one extra parameter (i.e., the segmentation threshold $D_{thr}$), which is easy to tune. Note that for reasonable and fair comparisons, for each of our segmentation-driven method, we set the parameters the same as the corresponding original method.

\begin{table}[t]\small
\centering
\setlength\tabcolsep{1pt}
\caption{Parameters of the mesh denoising methods. The tuned parameters are shown in the same order in Table~\ref{table:quantitative}.}
\label{table:tablepara}  
\scalebox{0.87}{
\begin{tabular}{|l|l|l|}
\hline
Methods   & \begin{tabular}[c]{@{}l@{}}Number of \\ Parameters\end{tabular} & Parameters Description                                                                                                                                                                                                                                                                                                                          \\ \hline
L0        & $3$         & \begin{tabular}[c]{@{}l@{}}$\beta_{max}$: maximum value of beta\\ $\alpha_0$: initial value for alpha\\ $\lambda$: weight for the $L_0$ term in the target function\end{tabular}                                                                                                                                                                      \\ \hline
BNF       & $3$         & \begin{tabular}[c]{@{}l@{}}$\sigma_{s}$: variance parameter for the spatial kernel\\ $n_{iter}$: number of iterations for normal update\\ $v_{iter}$: number of iterations for vertex update\end{tabular}                                                                                                                                            \\ \hline
UNF       & $3$         & \begin{tabular}[c]{@{}l@{}}$T$: threshold for controlling the averaging weights\\ $n_{iter}$: number of iterations for normal update\\ $v_{iter}$: number of iterations for vertex update\end{tabular}                                                                                                                                              \\ \hline
GNF       & $5$         & \begin{tabular}[c]{@{}l@{}}$r$: radius for finding a geometrical neighborhood\\ $\sigma_{s}$: variance parameter for the spatial kernel\\ $\sigma_{r}$: variance parameter for the range kernel\\ $n_{iter}$: number of iterations for normal update\\ $v_{iter}$: number of iterations for vertex update\end{tabular}                          \\ \hline
L1 & $3$         & \begin{tabular}[c]{@{}l@{}}$\sigma$: variance parameter for the spatial kernel\\ $n_{iter}$: number of iterations for normal update\\ $v_{iter}$: number of iterations for vertex update\end{tabular}                                                                                                                                                \\ \hline
HLO       & $1$         & $iter$: number of filtering iterations                                                                                                                                                                                                                                                                                                                  \\ \hline
TGV       & $3$         & \begin{tabular}[c]{@{}l@{}}$\alpha_{1}$: first-order parameter\\ $\alpha_{0}$: second-order parameter\\ $\beta$: fidelity term\end{tabular}                                                                                                                                                                                                            \\ \hline
Ours      & $1+X$       & \begin{tabular}[c]{@{}l@{}}$D_{thr}$: edge-based segmentation threshold
\\ $X$: the parameters of other methods\end{tabular} \\ \hline
\end{tabular}
}
\end{table}



\begin{table}[htbp]\small
\centering
\setlength\tabcolsep{1pt}
\caption{Quantitative comparisons in regards to MAE (in degrees) and $E_v(\times10^{-3})$, where the best results are marked in bold. }
\label{table:quantitative} 
\scalebox{0.87}{
\begin{tabular}{|l|l|l|l|l|}
\hline
Models                                                                                                                                                 & Methods                                                                                  & MAE                                                                                             & $E_v(\times10^{-3})$                                                                               & Parameters                                                                                                                                                   \\ \hline
\multirow{2}{*}{\begin{tabular}[c]{@{}l@{}}Cube\\ ($\sigma_n$ = 0.2$l_{e}$)\\ (Fig.~\ref{fig:cube02-quan})\\ $\left|V\right|$: 12,288\\ $\left|F\right|$: 6,146\\ $D_{thr}$ = 0.01\end{tabular}}            & \begin{tabular}[c]{@{}l@{}}L0\\ BNF\\ UNF\\ GNF\\ L1\\ HLO\\ TGV\end{tabular}            & \begin{tabular}[c]{@{}l@{}}2.692\\ 2.325\\ 0.460\\ 0.530\\ 0.519\\ 6.459\\ 0.294\end{tabular}     & \begin{tabular}[c]{@{}l@{}}8.960\\ 4.758\\ 1.229\\ 1.267\\ 1.303\\ 7.621\\ 1.163\end{tabular}               & \begin{tabular}[c]{@{}l@{}}(1000, 0.0033, 0.01)\\ (0.45, 80, 40)\\ (0.35, 20, 50)\\ (2, 1, 0.35, 20, 10)\\ (80, 40, 45)\\ (3)\\ (0.7, 0.2, 100)\end{tabular} \\ \cline{2-5} 
                                                                                                                                                       & \begin{tabular}[c]{@{}l@{}}Ours (BNF)\\ Ours (UNF)\\ Ours (GNF)\\ Ours (L1)\end{tabular} & \begin{tabular}[c]{@{}l@{}}\textbf{0.276}\\ 0.447\\ 0.436\\ 0.486\end{tabular}                             & \begin{tabular}[c]{@{}l@{}}\textbf{0.942}\\ 1.197\\ 1.207\\ 1.275\end{tabular}                                       & \begin{tabular}[c]{@{}l@{}}(0.45, 80, 40)\\ (0.35, 20, 50)\\ (2, 1, 0.35, 20, 10)\\ (80, 40, 45)\end{tabular}                                                \\ \hline
\multirow{2}{*}{\begin{tabular}[c]{@{}l@{}}Casting \\ ($\sigma_n$ = 0.1$l_{e}$)\\ (Fig.~\ref{fig:casting01-quan})\\ $\left|V\right|$: 5,086\\ $\left|F\right|$: 10,204\\ $D_{thr}$ = 0.001\end{tabular}}       & \begin{tabular}[c]{@{}l@{}}L0\\ BNF\\ UNF\\ GNF\\ L1\\ HLO\\ TGV\end{tabular}            & \begin{tabular}[c]{@{}l@{}}9.386\\ 8.039\\ 12.109\\ 10.685\\ 6.642\\ 13.213\\ \textbf{3.514}\end{tabular}  & \begin{tabular}[c]{@{}l@{}}2.452\\ 2.244\\ 4.970\\ 1.682\\ \textbf{0.864}\\ 4.487\\ 0.970\end{tabular}               & \begin{tabular}[c]{@{}l@{}}(1000, 0.0051, 0.0001)\\ (0.35, 20, 10)\\ (0.4, 20, 20)\\ (2, 1, 0.25, 5, 5)\\ (20, 10, 30)\\ (3)\\ (0.7, 0.05, 100)\end{tabular}   \\ \cline{2-5} 
                                                                                                                                                       & \begin{tabular}[c]{@{}l@{}}Ours (BNF)\\ Ours (UNF)\\ Ours (GNF)\\ Ours (L1)\end{tabular} & \begin{tabular}[c]{@{}l@{}}7.731\\ 12.486\\ 7.189\\ 6.633\end{tabular}                            & \begin{tabular}[c]{@{}l@{}}2.059\\ 4.640\\ 1.191\\ \textbf{0.864}\end{tabular}                                       & \begin{tabular}[c]{@{}l@{}}(0.35, 20, 10)\\ (0.4, 20, 20)\\ (2, 1, 0.25, 5, 5)\\ (20, 10, 30)\end{tabular}                                                   \\ \hline
\multirow{2}{*}{\begin{tabular}[c]{@{}l@{}}Double Torus \\ ($\sigma_n$ = 0.2$l_{e}$)\\ (Fig.~\ref{fig:dbltorus02-quan})\\ $\left|V\right|$: 8,702\\ $\left|F\right|$: 17,408\\ $D_{thr}$ = 0.01\end{tabular}}   & \begin{tabular}[c]{@{}l@{}}L0\\ BNF\\ UNF\\ GNF\\ L1\\ HLO\\ TGV\end{tabular}            & \begin{tabular}[c]{@{}l@{}}\textbf{1.109}\\ 8.049\\ 2.869\\ 2.697\\ 3.116\\ 8.781\\ 1.181\end{tabular}     & \begin{tabular}[c]{@{}l@{}}7.367\\ 50.788\\ 8.509\\ 10.126\\ 9.718\\ 50.171\\ 8.132\end{tabular}            & \begin{tabular}[c]{@{}l@{}}(1000, 0.0024, 0.01)\\ (0.45, 80, 40)\\ (0.35, 20, 50)\\ (2, 1, 0.35, 20, 10)\\ (80, 40, 45)\\ (3)\\ (0.7, 0.2, 100)\end{tabular} \\ \cline{2-5} 
                                                                                                                                                       & \begin{tabular}[c]{@{}l@{}}Ours (BNF)\\ Ours (UNF)\\ Ours (GNF)\\ Ours (L1)\end{tabular} & \begin{tabular}[c]{@{}l@{}}2.520\\ 2.832\\ 2.403\\ 3.034\end{tabular}                             & \begin{tabular}[c]{@{}l@{}}9.943\\ 8.282\\ \textbf{7.207}\\ 9.735\end{tabular}                                       & \begin{tabular}[c]{@{}l@{}}(0.45, 80, 40)\\ (0.35, 20, 50)\\ (2, 1, 0.35, 20, 10)\\ (80, 40, 45)\end{tabular}                                                \\ \hline
\multirow{2}{*}{\begin{tabular}[c]{@{}l@{}}Dodecahedron\\ ($\sigma_n$ = 0.4$l_{e}$)\\ (Fig.~\ref{fig:dodecahedron04-quan})\\ $\left|V\right|$: 4,610\\ $\left|F\right|$: 9,216\\ $D_{thr}$ = 0.005\end{tabular}}   & \begin{tabular}[c]{@{}l@{}}L0\\ BNF\\ UNF\\ GNF\\ L1\\ HLO\\ TGV\end{tabular}            & \begin{tabular}[c]{@{}l@{}}12.549\\ 15.350\\ 14.926\\ 8.833\\ 13.846\\ 9.641\\ 8.967\end{tabular} & \begin{tabular}[c]{@{}l@{}}25.771\\ 13.808\\ 11.117\\ 6.832\\ 8.213\\ 16.432\\ 8.219\end{tabular}           & \begin{tabular}[c]{@{}l@{}}(1000, 0.0038, 5)\\ (0.33, 40, 20)\\ (0.4, 15, 10)\\ (2, 1, 0.2, 75, 20)\\ (80, 40, 45)\\ (5)\\ (0.7. 0.2, 100)\end{tabular}      \\ \cline{2-5} 
                                                                                                                                                       & \begin{tabular}[c]{@{}l@{}}Ours (BNF)\\ Ours (UNF)\\ Ours (GNF)\\ Ours (L1)\end{tabular} & \begin{tabular}[c]{@{}l@{}}8.976\\ 10.366\\ \textbf{7.159}\\ 10.449\end{tabular}                           & \begin{tabular}[c]{@{}l@{}}3.878\\ 5.007\\ \textbf{2.695}\\ 5.486\end{tabular}                                       & \begin{tabular}[c]{@{}l@{}}(0.33, 40, 20)\\ (0.4, 15, 10)\\ (2, 1, 0.2, 75, 20)\\ (80, 40, 45)\end{tabular}                                                  \\ \hline
\multirow{2}{*}{\begin{tabular}[c]{@{}l@{}}Icosahedron\\ ($\sigma_n$ = 0.4$l_{e}$)\\ (Fig.~\ref{fig:icosahedron04-quan})\\ $\left|V\right|$: 10,242\\ $\left|F\right|$: 20,480\\ $D_{thr}$ = 0.0008\end{tabular}} & \begin{tabular}[c]{@{}l@{}}L0\\ BNF\\ UNF\\ GNF\\ L1\\ HLO\\ TGV\end{tabular}            & \begin{tabular}[c]{@{}l@{}}9.440\\ 9.859\\ 8.361\\ 3.534\\ 5.925\\ 6.472\\ 3.576\end{tabular}     & \begin{tabular}[c]{@{}l@{}}5.604\\ 3.380\\ 2.057\\ 1.155\\ 1.463\\ 2.909\\ 3.391\end{tabular}               & \begin{tabular}[c]{@{}l@{}}(1000, 0.0022, 1)\\ (0.33, 40, 20)\\ (0.4, 15, 10)\\ (2, 1, 0.2, 75, 20)\\ (80, 40, 45)\\ (5)\\ (0.7, 0.2, 100)\end{tabular}      \\ \cline{2-5} 
                                                                                                                                                       & \begin{tabular}[c]{@{}l@{}}Ours (BNF)\\ Ours (UNF)\\ Ours (GNF)\\ Ours (L1)\end{tabular} & \begin{tabular}[c]{@{}l@{}}3.712\\ 4.685\\ \textbf{3.111}\\ 4.237\end{tabular}                             & \begin{tabular}[c]{@{}l@{}}0.946\\ 1.054\\ \textbf{0.921}\\ 1.002\end{tabular}                                       & \begin{tabular}[c]{@{}l@{}}(0.33, 40, 20)\\ (0.4, 15, 10)\\ (2, 1, 0.2, 75, 20)\\ (80, 40, 45)\end{tabular}                                                  \\ \hline
\multirow{2}{*}{\begin{tabular}[c]{@{}l@{}}Cad\\ ($\sigma_n$ = 0.3$l_{e}$)\\ (Fig.~\ref{fig:cad03-quan})\\ $\left|V\right|$: 19,398\\ $\left|F\right|$: 38,792\\ $D_{thr}$ = 0.001\end{tabular}}          & \begin{tabular}[c]{@{}l@{}}L0\\ BNF\\ UNF\\ GNF\\ L1\\ HLO\\ TGV\end{tabular}            & \begin{tabular}[c]{@{}l@{}}2.673\\ 2.418\\ 2.822\\ 2.730\\ 2.807\\ 11.223\\ 2.029\end{tabular}    & \begin{tabular}[c]{@{}l@{}}202.264\\ 207.254\\ 239.814\\ 252.715\\ 186.496\\ 793.123\\ 377.546\end{tabular} & \begin{tabular}[c]{@{}l@{}}(1000, 0.0013, 1)\\ (0.35, 25, 20)\\ (0.55, 20, 40)\\ (2, 1, 0.25, 25, 20)\\ (80, 40, 45)\\ (5)\\ (0.7. 0.2, 100)\end{tabular}    \\ \cline{2-5} 
                                                                                                                                                       & \begin{tabular}[c]{@{}l@{}}Ours (BNF)\\ Ours (UNF)\\ Ours (GNF)\\ Ours (L1)\end{tabular} & \begin{tabular}[c]{@{}l@{}}\textbf{2.021}\\ 2.523\\ 2.439\\ 2.486\end{tabular}                             & \begin{tabular}[c]{@{}l@{}}181.763\\ 219.908\\ 245.837\\ \textbf{165.722}\end{tabular}                               & \begin{tabular}[c]{@{}l@{}}(0.35, 25, 20)\\ (0.55, 20, 40)\\ (2, 1, 0.25, 25, 20)\\ (80, 40, 45)\end{tabular}                                                \\ \hline
\end{tabular}
}
\end{table}

\subsection{Quantitative Evaluations}
We firstly show quantitative evaluations on mesh models with synthetic Gaussian noise. Such noisy mesh models are generated by adding zero-mean Gaussian noise with standard deviation $\sigma_n$ to the corresponding ground truth models. Here, $\sigma_n$ describes the noise level, which is proportional to the mean edge length $l_{e}$ of the input mesh, with the parameters shown in Table~\ref{table:quantitative}. As the corresponding clean mesh models are known, we employ two common metrics as suggested by previous works \cite{sun2007fast, Zheng2011, liu_tgv_2021} to respectively evaluate the errors on denoised face normals and vertex positions: the Mean Angular Error (MAE) metric, which measures the average angular error between the face normals on the filtered mesh and the ones of the ground truth mesh; and a vertex-based mesh-to-mesh error metric ($E_v$), which measures the accuracy of the denoised vertices' positions.

\begin{figure*}[htb!]
	\centering{
		\includegraphics[width=\textwidth]{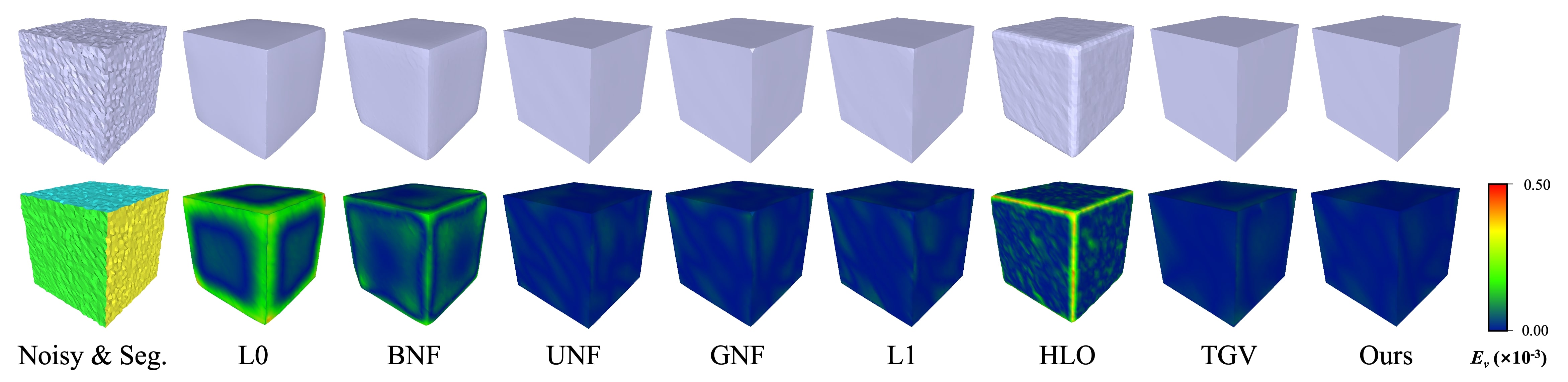}
	}
	\caption{Coloured $E_v$ for Cube with noise $\sigma_{n}=0.2l_e$, with \cite{Lee2005} as our denoise backbone.} 
	\label{fig:cube02-quan}
\end{figure*}

\begin{figure*}[htb!]
	\centering{
		\includegraphics[width=\textwidth]{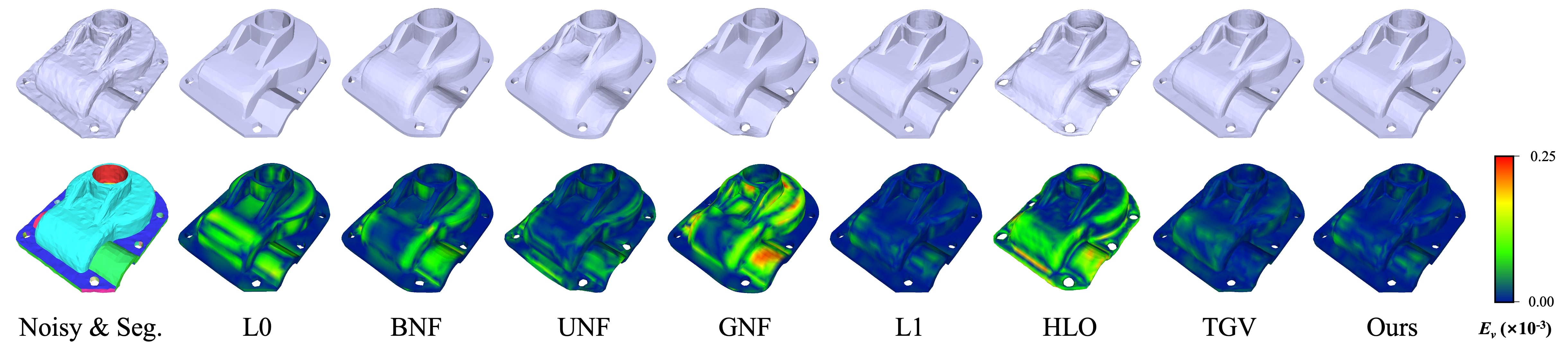}
	}
	\caption{Coloured $E_v$ for Casting with noise $\sigma_{n}=0.1l_e$, with \cite{Lu2017-L1median} as our denoise backbone.} 
	\label{fig:casting01-quan}
\end{figure*}

\begin{figure*}[htb!]
	\centering{
		\includegraphics[width=\textwidth]{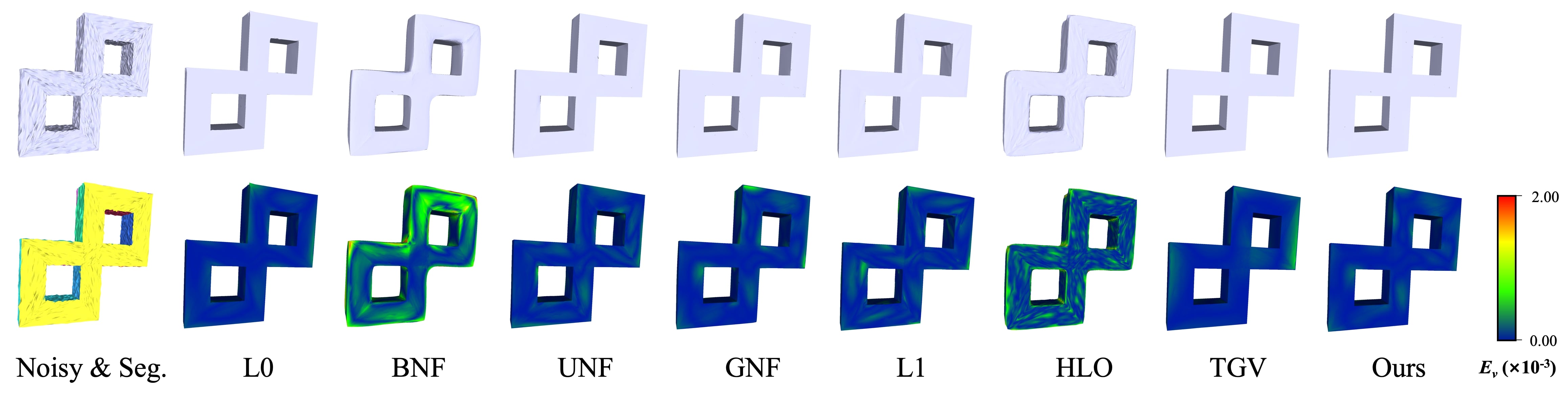}
	}
	\caption{Coloured $E_v$ for Double Torus with noise $\sigma_{n}=0.2l_e$, with \cite{zhang2015guided} as our denoise backbone.} 
	\label{fig:dbltorus02-quan}
\end{figure*}

\begin{figure*}[htb!]
	\centering{
		\includegraphics[width=\textwidth]{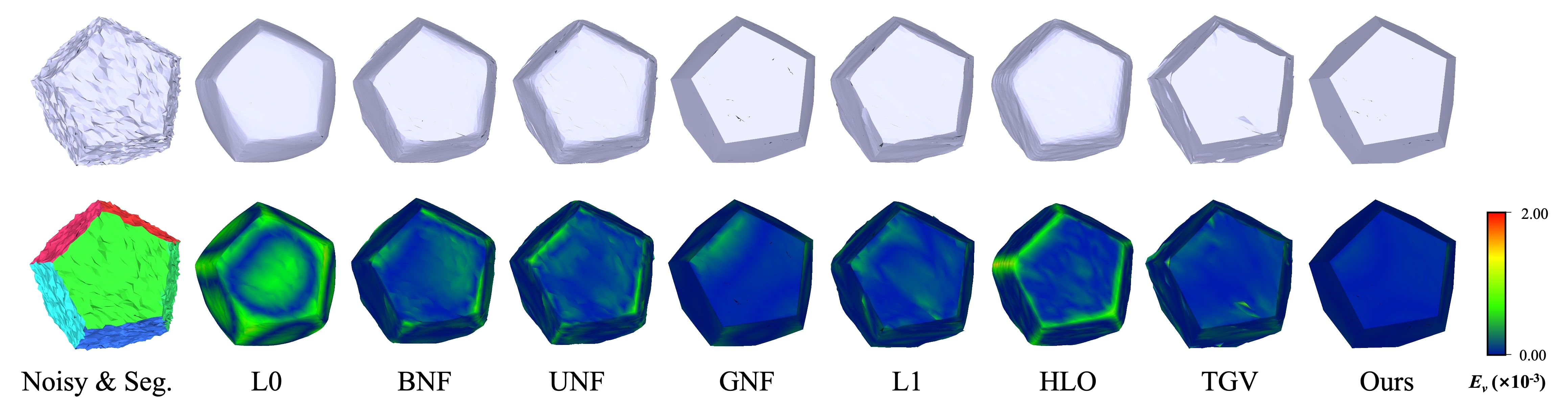}
	}
	\caption{Coloured $E_v$ for Dodecahedron with noise $\sigma_{n}=0.4l_e$, with \cite{zhang2015guided} as our denoise backbone.} 
	\label{fig:dodecahedron04-quan}
\end{figure*}

\begin{figure*}[htb!]
	\centering{
		\includegraphics[width=\textwidth]{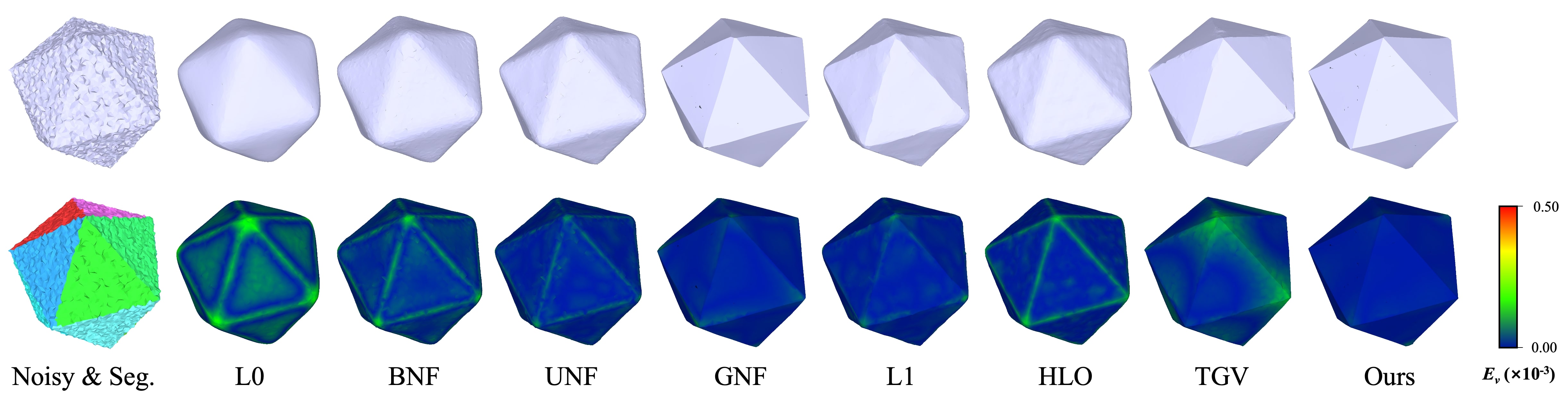}
	}
	\caption{Coloured $E_v$ for Icosahedron with noise $\sigma_{n}=0.4l_e$, with \cite{zhang2015guided} as our denoise backbone.} 
	\label{fig:icosahedron04-quan}
\end{figure*}

\begin{figure*}[htb!]
	\centering{
		\includegraphics[width=\textwidth]{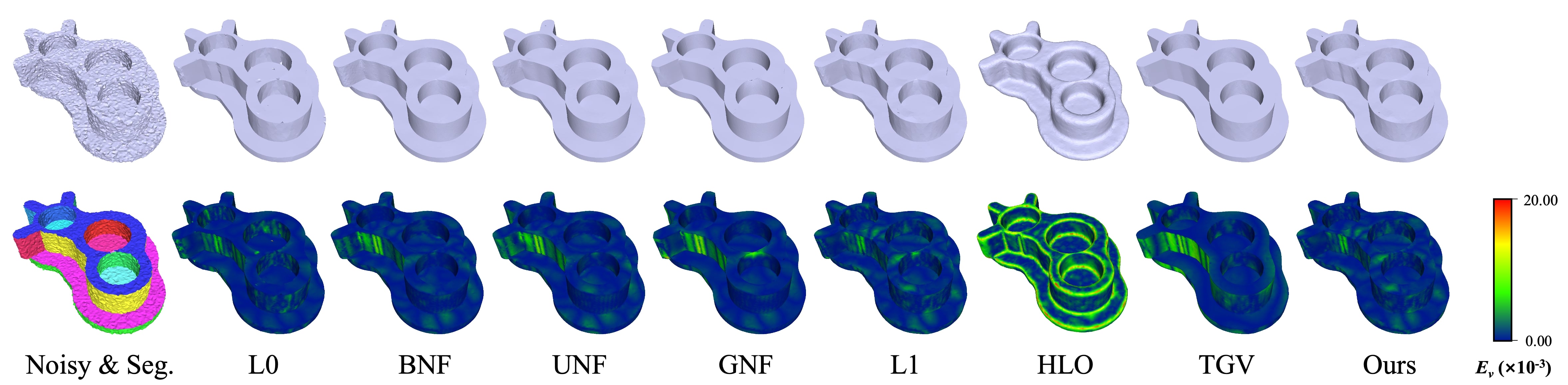}
	}
	\caption{Coloured $E_v$ for Cad with noise $\sigma_{n}=0.3l_e$, with \cite{Lu2017-L1median} as our denoise backbone.} 
	\label{fig:cad03-quan}
\end{figure*}

Table~\ref{table:quantitative} lists MAE and $E_v$ over six models for all methods. Our segmentation-driven backbone achieves the minimum $E_v$ metric on all shapes, and achieves competitive results for MAE. To demonstrate our advantages, we visualise the $E_v$ values with colour gradients from Fig.~\ref{fig:cube02-quan} to Fig.~\ref{fig:cad03-quan}, along with our segmentation results (abbreviated as Seg. in the figures). 
Our advantages in feature preservation can be especially seen in Fig.~\ref{fig:cube02-quan}, \ref{fig:dodecahedron04-quan} and \ref{fig:icosahedron04-quan}: despite L0 and BNF are generally good at dealing with CAD-like shapes, they tend to blur sharp edges on such shapes (which are reflected by the green colour near the edges); TGV generally performs well, but may still not be able to robustly preserve the edges; by contrast, with the assistance from our segmentation backbone, the $E_v$ error on such shapes denoised by our method (especially on the edges) is significantly reduced.

\begin{figure*}[!t]
\centering {
    \begin{minipage}[t]{0.45\textwidth}
    \centering
    \includegraphics[width=\columnwidth]{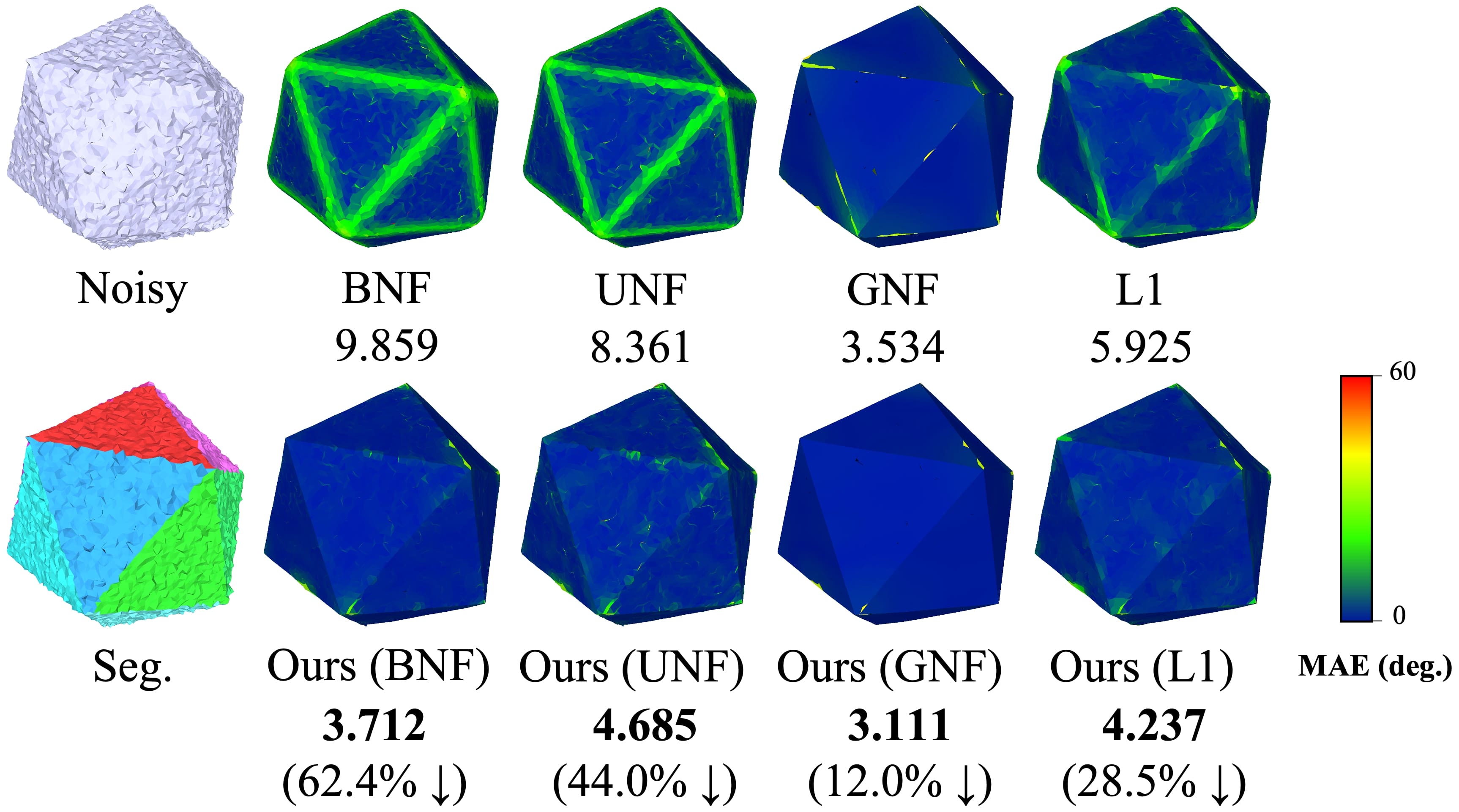}
    
    \end{minipage}
    \begin{minipage}[t]{0.45\textwidth}
    \centering
    \includegraphics[width=\columnwidth]{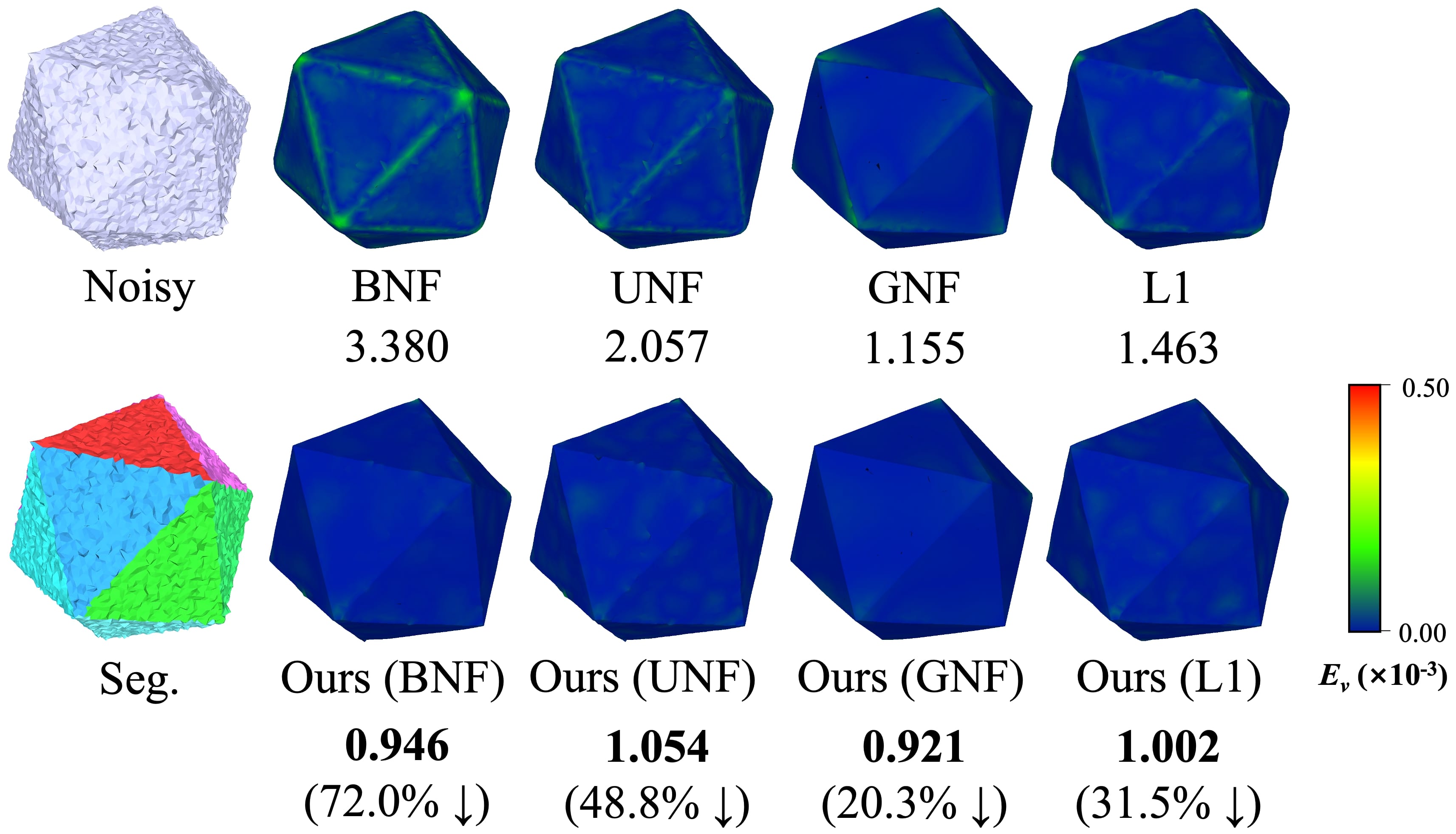}
    \end{minipage}
}

\caption{The MAE values (left, in degrees) and the $E_v$ values (right, in $\times10^{-3}$), along with the changes in percentages after embedding our segmentation backbone.}
\label{fig:icosahedron-improve}

\end{figure*}

In addition, as demonstrated in Table~\ref{table:quantitative}, our method significantly boosts the denoising effects of each original method. Fig.~\ref{fig:icosahedron-improve} demonstrate the improvements on the MAE and $E_v$ metrics respectively on the denoised Icosahedron model (the fifth shape in Table~\ref{table:quantitative}), showing the metric values and the changes in percentages.


\subsection{Visual Results}

\textit{Synthetic Models.} We first compare visual results on various synthetic models corrupted with Gaussian noise. Analogous to previous research \cite{sun2007fast,Lu2016}, we find that the visual comparisons might be inconsistent with MAE and $E_v$. Thus, we focus on the visual effects in this section.

\begin{figure*}[htb!]
	\centering{
		\includegraphics[width=\textwidth]{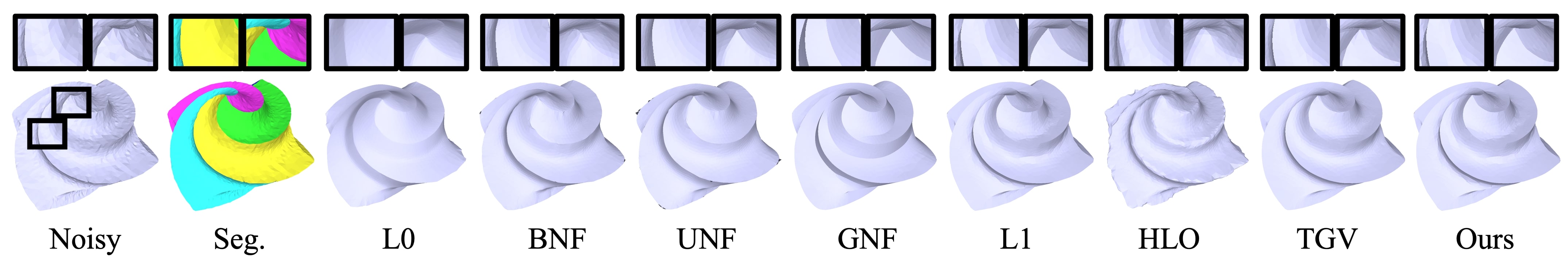}
	}
	\caption{Visual comparison for Octaflower with noise $\sigma_{n}=0.1l_e$, with \cite{Lu2017-L1median} as our denoise backbone.} 
	\label{fig:octaflower01-vis}
\end{figure*}

\begin{figure*}[htb!]
	\centering{
		\includegraphics[width=\textwidth]{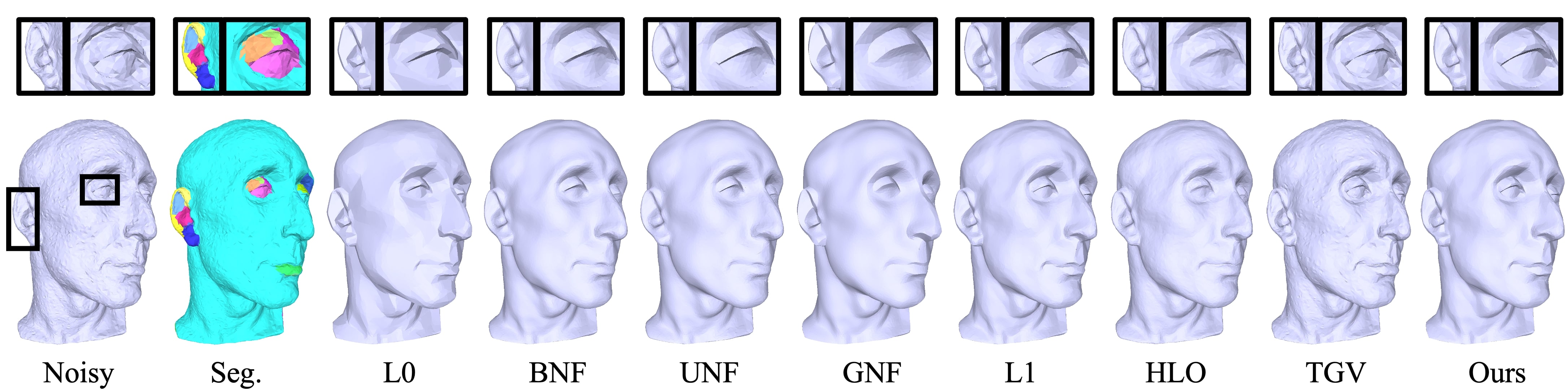}
	}
	\caption{Visual comparison for Nicolo with noise $\sigma_{n}=0.1l_e$, with \cite{Lu2017-L1median} as our denoise backbone.} 
	\label{fig:nicolo01-vis}
\end{figure*}

\begin{figure*}[htb!]
	\centering{
		\includegraphics[width=\textwidth]{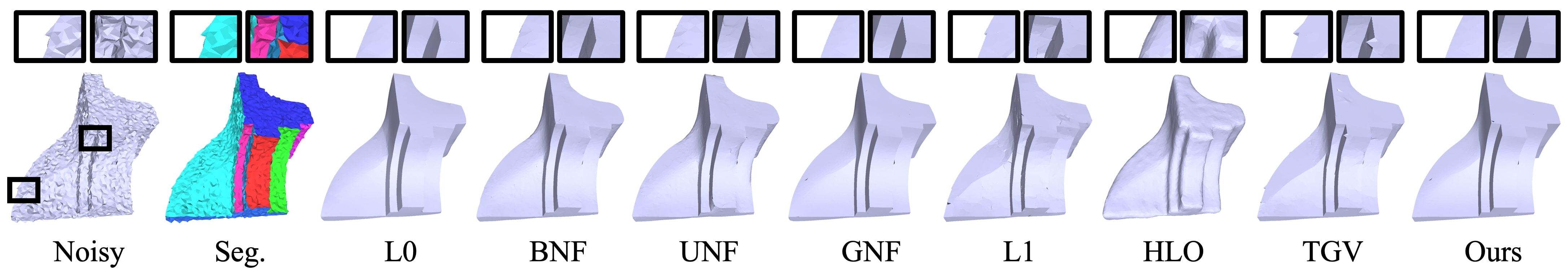}
	}
	\caption{Visual comparison for Fandisk with noise $\sigma_{n}=0.4l_e$, with \cite{zhang2015guided} as our denoise backbone.} 
	\label{fig:fandisk04-vis}
\end{figure*}

As shown in Fig.~\ref{fig:octaflower01-vis}, our segmentation-driven approach assists with preserving the sharp tip on the Octaflower mesh while preserving the curve regions. Although the tip is blurred in the noisy shape, our segmentation method robustly recognises it and helps the denoising algorithm to reveal it. Also, unlike \cite{zhang2015guided} which oversharpens the curved regions, our method keeps the triangular facets on each curve surfaces within the same region, which assists in maintaining the curvatures of such areas. Similarly, as displayed in Fig.~\ref{fig:nicolo01-vis}, our segmentation approach helps with restoring the details of the ear and the eye on the Nicolo model. Furthermore, most anisotropic methods (such as \cite{Hildebrandt2004}) often struggle with preserving features at higher noise levels (i.e., $\sigma_n\ge0.3l_{e}$) \cite{Lu2017-L1median}. This can also be seen from Fig.~\ref{fig:fandisk04-vis}, where BNF \cite{Lee2005} cannot fully preserve features of the Fandisk model that is corrupted with severe noise. In contrast, our segmentation-driven method can produce smooth surfaces and preserve the sharp features on this model.


\begin{figure*}[htb!]
	\centering{
		\includegraphics[width=\textwidth]{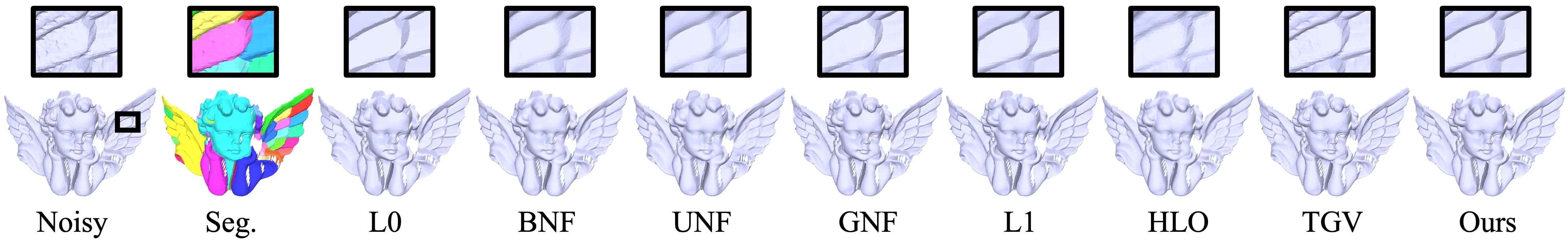}
	}
	\caption{Visual comparison for a raw scanned Angel mesh, with \cite{Lu2017-L1median} as our denoise backbone.} 
	\label{fig:angel-real-vis}
\end{figure*}

\begin{figure*}[htb!]
	\centering{
		\includegraphics[width=\textwidth]{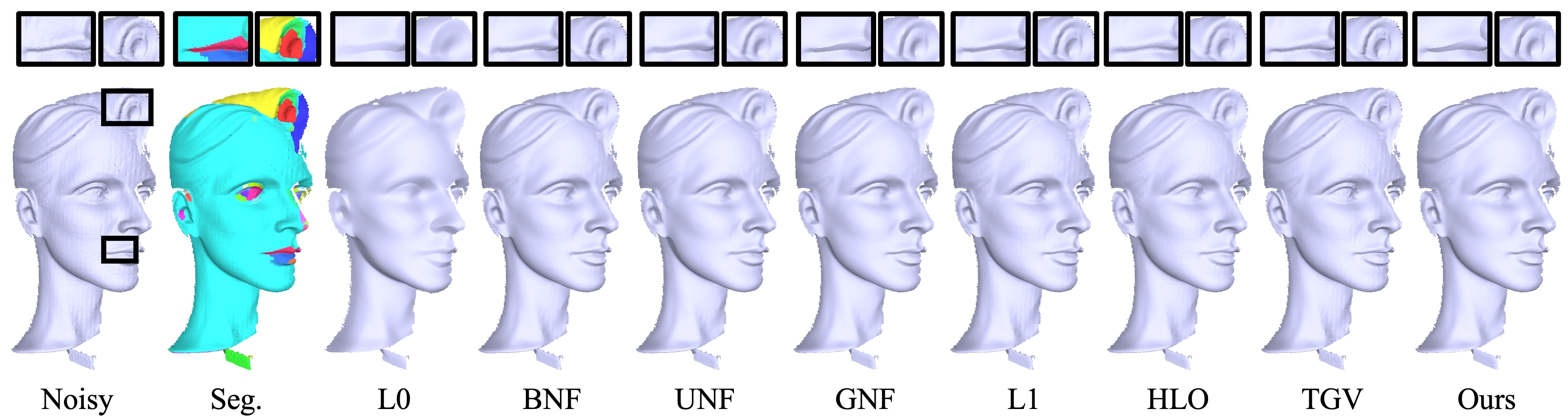}
	}
	\caption{Visual comparison for a raw scanned Wilhelm mesh, with \cite{zhang2015guided} as our denoise backbone.} 
	\label{fig:wilhelm-real-vis}
\end{figure*}

\begin{figure*}[htb!]
	\centering{
		\includegraphics[width=\textwidth]{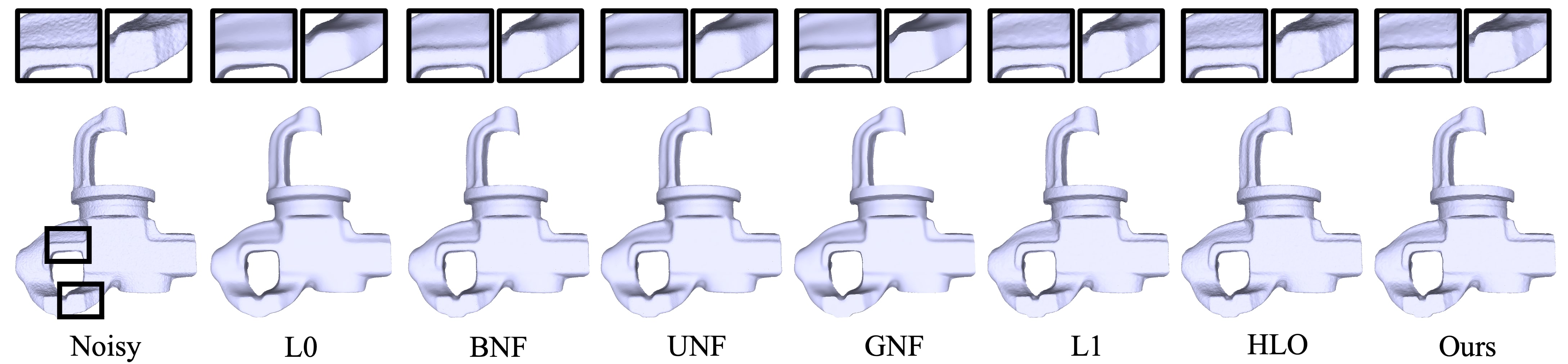}
	}
	\caption{Visual comparison for a raw scanned Iron mesh, with \cite{Lu2017-L1median} as our denoise backbone. } 
	\label{fig:iron-real-vis}
\end{figure*}

\begin{figure*}[htb!]
	\centering{
		\includegraphics[width=\textwidth]{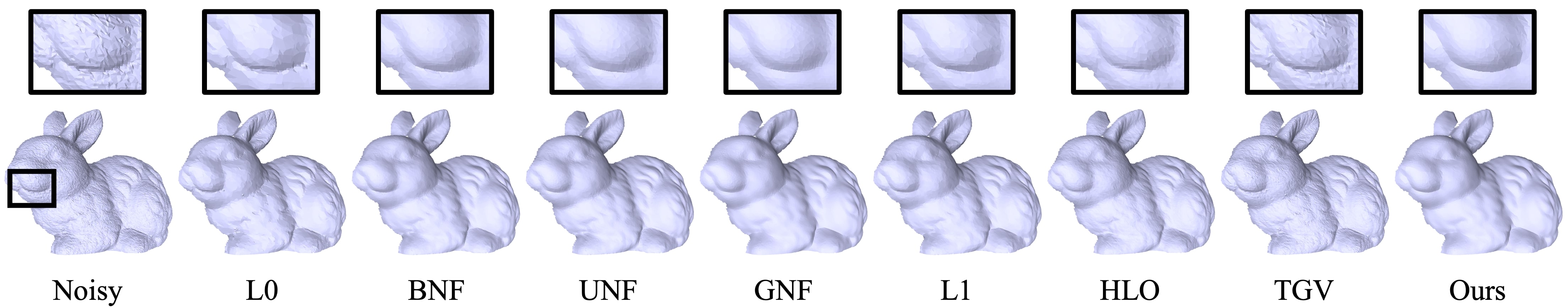}
	}
	\caption{Visual comparison for a raw scanned Rabbit mesh, with \cite{zhang2015guided} as our denoise backbone. } 
	\label{fig:rabbit-real-vis}
\end{figure*}

\textit{Raw Scanned Models.} In addition to the synthetic shapes, we compare denoising performance on scanned models corrupted with raw noise. Such shapes come from scanners and their noise is different from Gaussian noise. Despite the fact, our segmentation-driven pipeline can still preserve original features and details of the input meshes. For example, in Fig.~\ref{fig:angel-real-vis}, the details of the feathers on the Angel model are preserved. Similarly, on the Wilhelm model in Fig.~\ref{fig:wilhelm-real-vis}, the details on the hair and the lips are maintained. Besides, our segmentation approach can also help with restoring details that are corrupted with the scanner's noise, such as the Iron model in Fig.~\ref{fig:iron-real-vis} and the Rabbit model in Fig.~\ref{fig:rabbit-real-vis}.

\section{Discussion}
\label{sec:discussion}

\subsection{Segmentation Methods}
\label{subsec:segmentation-methods}

We attempted various mesh segmentation backbones: \textit{K-means segmentation}, a classic method that has been adopted throughout the years \cite{lian_adaptive_2022}; \textit{region-growing segmentation} (abbreviated as R.G.) \cite{yang-regiongrowth-2014, benzian-regiongrowth-2020}, where we adopt $N_{thr}$ as our boundary-detection metric; \textit{semantic segmentation}, where we adopt a commonly-used algorithm based on Shape Diameter Function (abbreviated as SDF) \cite{shapira2008consistent} that is also deployed in the CGAL library \cite{cgal}; and finally, our \textit{edge-based segmentation} strategy.

\begin{figure}[t]
	\centering{
		\includegraphics[width=\columnwidth]{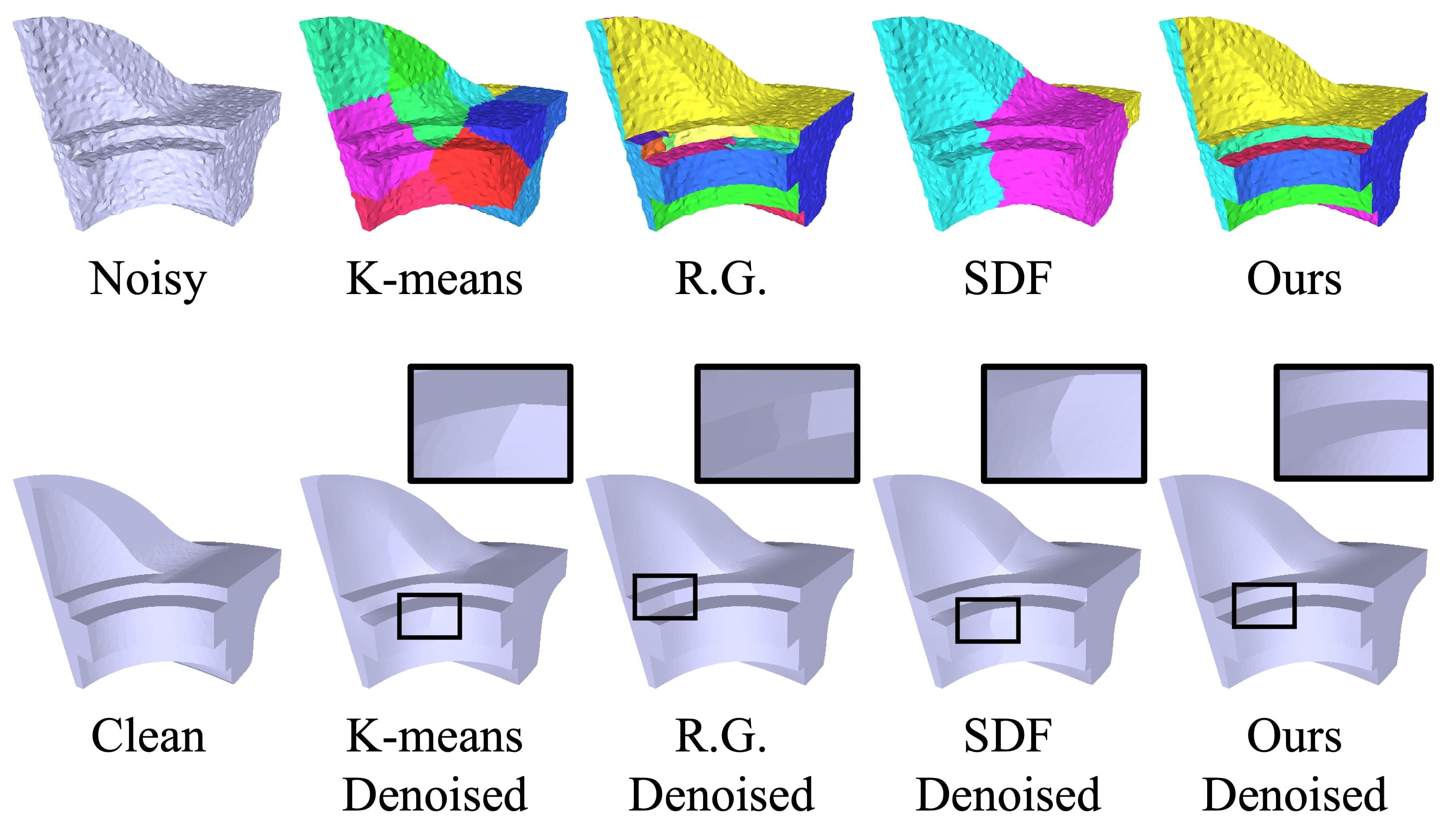}
	}
	\caption{Different segmentation techniques and their results on segmentation and denoising on a Fandisk model with $\sigma_n=0.2l_e$.} 
	\label{fig:segmentation_compare}
\end{figure}

Fig.~\ref{fig:segmentation_compare} shows the segmentation results of these methods on a Fandisk mesh model with $\sigma_n=0.2l_e$. The denoising approach is \cite{zhang2015guided}, where the tuned parameters are $(r, \sigma_{s}, \sigma_{r}, n_{iter}, v_{iter})=(2, 1, 0.25, 25, 20)$ and $D_{thr} = 0.02$. For the K-means algorithm, we manually set $K=12$ as such a setting is optimised for the geometry of Fandisk; we also set the clustering iterations to 5 to obtain more robust results. As can be seen from the denoised results, the segmentation outcome of K-means is not informative at all; region-growing method is fragile to noise, resulting in sub-optimal segments; as \cite{shapira2008consistent} is semantic segmentation, it fails to segment the shape based on geometric information. As a result, the above three segmentation methods all give false boundaries, which result in unwanted folds in the denoised meshes. By contrast, our method accurately clusters all smooth regions and the denoised result is closer to the ground truth.

\begin{figure}[t]
	\centering{
		\includegraphics[width=0.8\columnwidth]{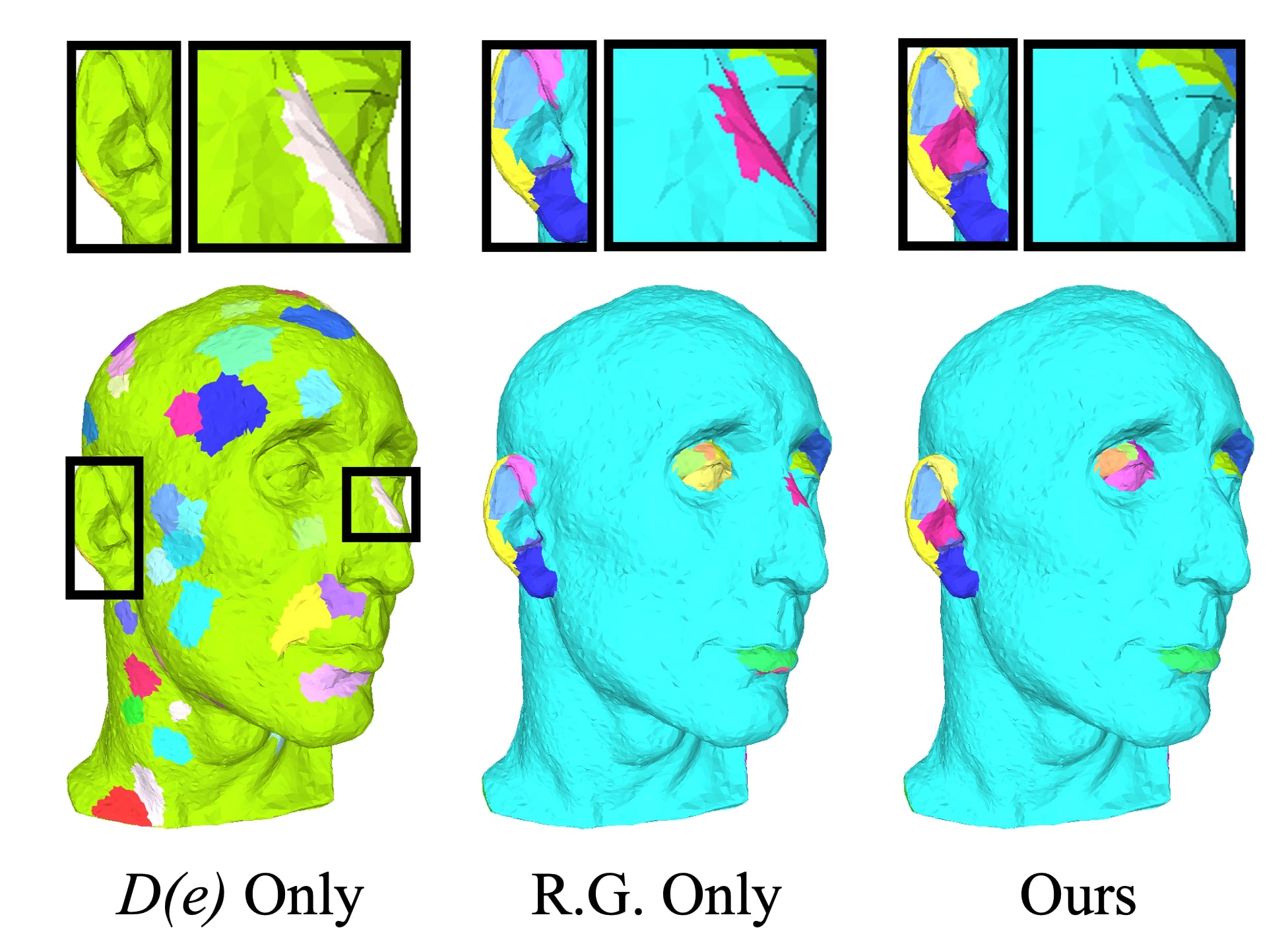}
	}
	\caption{Different strategies for segmentation on a Nicolo model with noise $\sigma_{n}=0.1l_{e}$, where our segmentation strategy provides the most informative segmentation result.}
	\label{fig:strategy-comparison}
\end{figure}

We also compare for different segmentation strategies. The segmentation results in Fig.~\ref{fig:strategy-comparison} demonstrate that only utilising $N_{thr}$ or $D(e)$ is not enough. Merely relying on $D(e)$ does not give informative regions, resulting in a large region with small patches. Only using $N_{thr}$ for region growing may result in unwanted clusters, such as the extra magenta area on the nose, and some detailed regions (e.g., earhole) are not well-segmented. By contrast, our combined approach provides a much more reasonable segmentation result that facilitates denoising.

\subsection{Algorithms Stability}
\label{subsec:algorithm-stability}
\begin{figure*}[t]
	\centering{
		\includegraphics[width=0.95\textwidth]{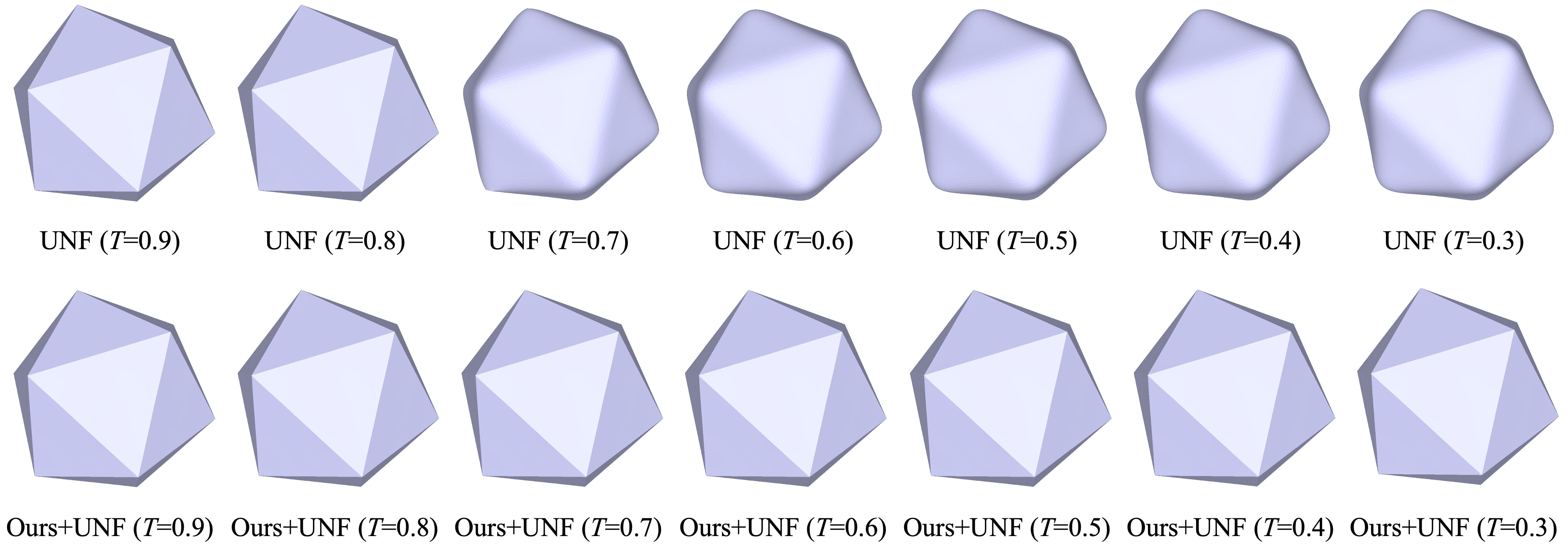}
	}
	\caption{Stability on denoising Icosahedron ($\sigma_n=0.1l_e$) with different $T$ using UNF \cite{sun2007fast}.} 
	\label{fig:unf-stability}
\end{figure*}

\begin{figure*}[htbp]
\centering {
\begin{minipage}[t]{0.48\textwidth}
\centering
\includegraphics[width=0.9\columnwidth]{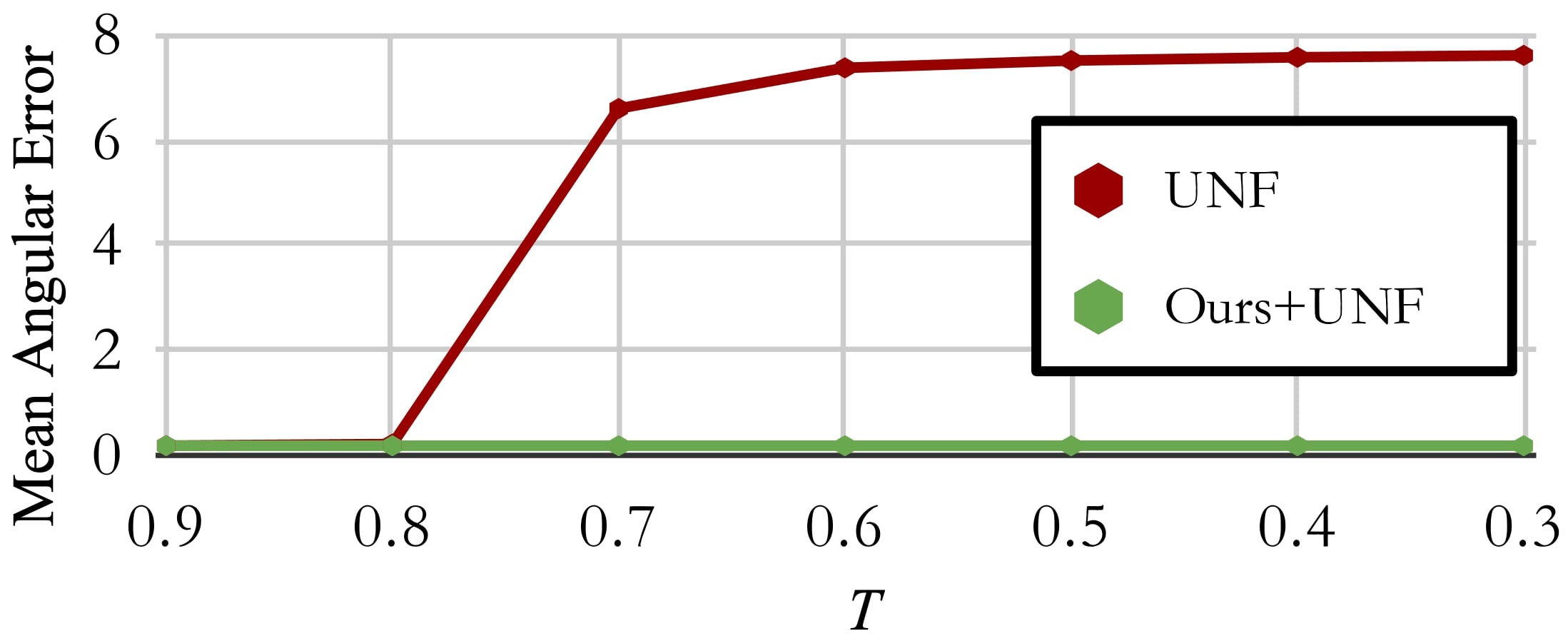}
\end{minipage}
\begin{minipage}[t]{0.48\textwidth}
\centering
\includegraphics[width=0.9\columnwidth]{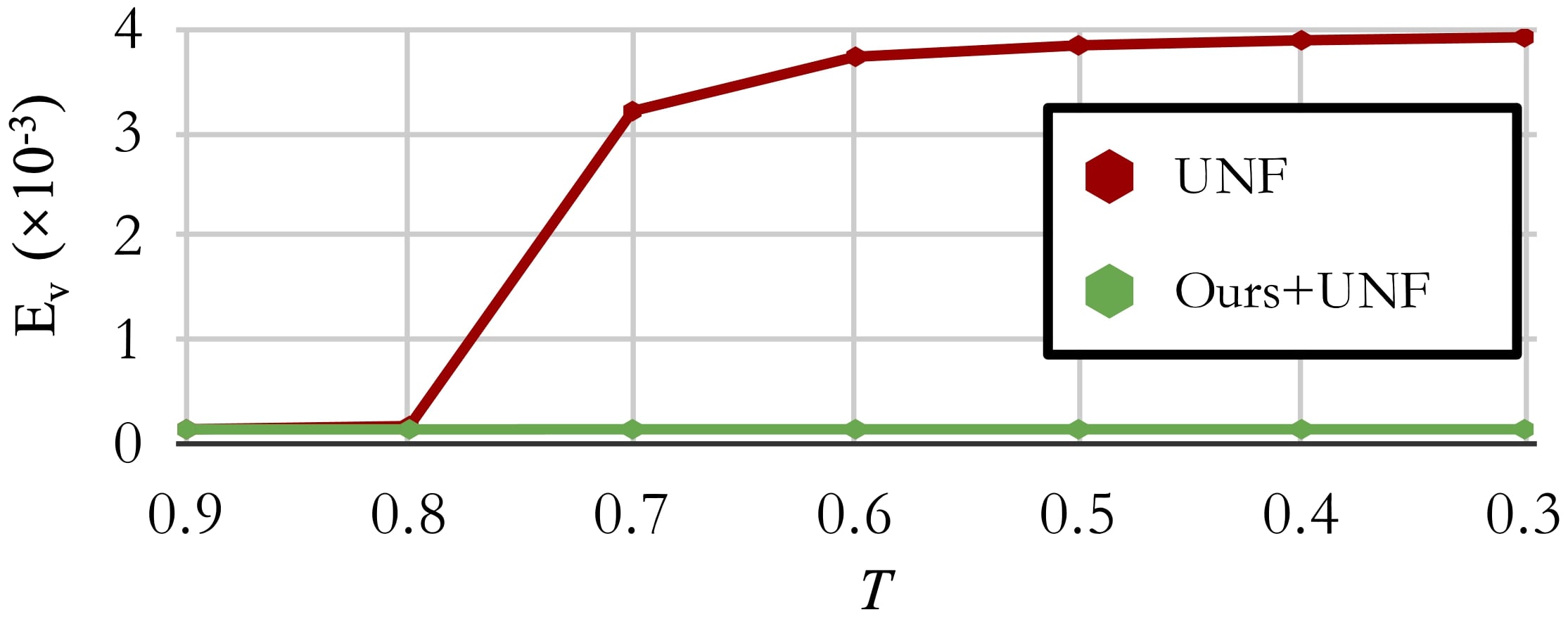}
\end{minipage}
}

\caption{The changes in MAE (left) and $E_v$ (right) with different $T$ values. }
\label{fig:unf-stability-trend}

\end{figure*}

Our segmentation result also increases the stability of the original algorithms, making them more robust against parameter variations. For example, Fig.~\ref{fig:unf-stability} shows the denoising results of an Icosahedron model using the original version of UNF and the version with our segmentation backbone, tested with different values of threshold $T$. For the original UNF method, the visual results are quite sensitive to different $T$ values. After embedding our backbone, the sharp edges are generally well-preserved, even with different $T$. This can also be seen from the line charts of the computed MAE and $E_v$ values in Fig.~\ref{fig:unf-stability-trend}.

\begin{figure*}[t]
	\centering{
		\includegraphics[width=0.95\textwidth]{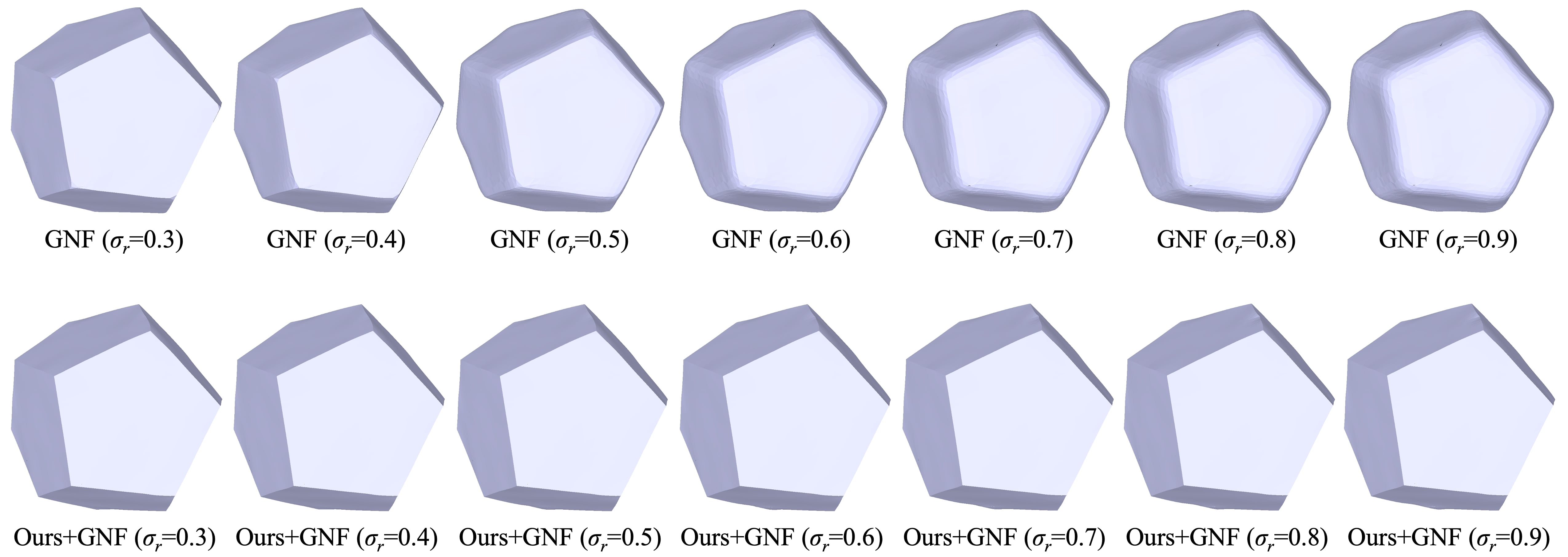}
	}
	\caption{Stability on denoising Dodecahedron ($\sigma_n=0.2l_e$) with different $\sigma_r$ using GNF \cite{zhang2015guided}.} 
	\label{fig:gnf-stability}
\end{figure*}

\begin{figure*}[htbp]
\centering {
\begin{minipage}[t]{0.48\textwidth}
\centering
\includegraphics[width=0.9\columnwidth]{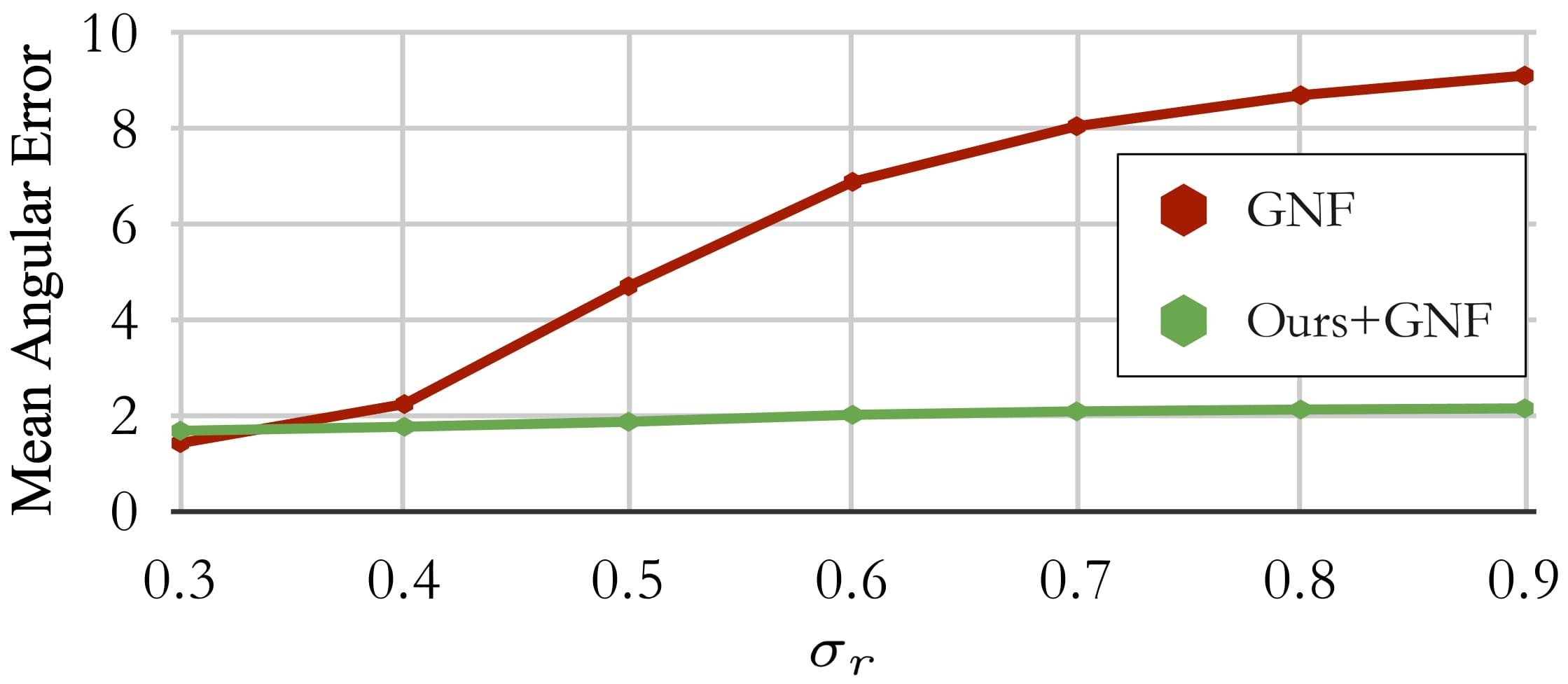}
\end{minipage}
\begin{minipage}[t]{0.48\textwidth}
\centering
\includegraphics[width=0.9\columnwidth]{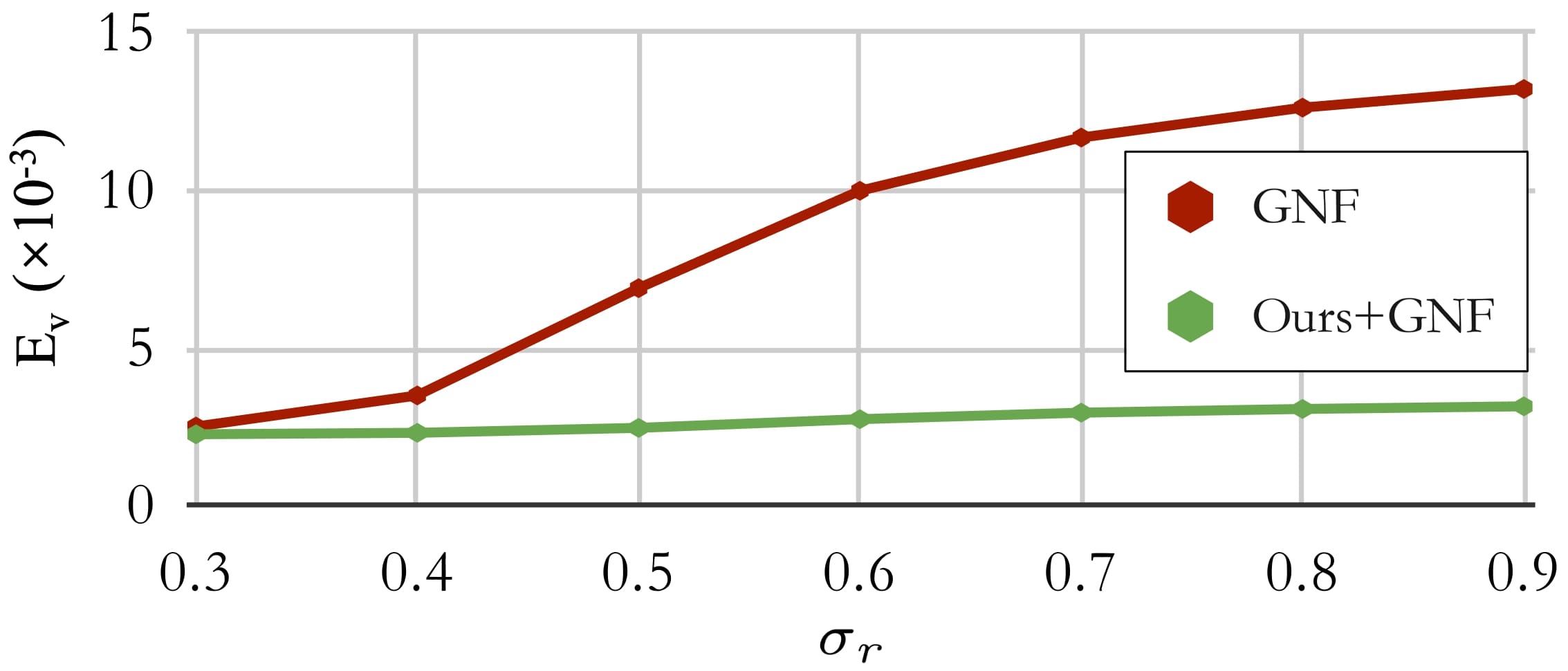}
\end{minipage}
}
\caption{The changes in MAE (left) and $E_v$ (right) with different $\sigma_r$ values.}
\label{fig:gnf-stability-trend}

\end{figure*}

Similarly, Fig.~\ref{fig:gnf-stability} shows the denoising results on a Dodecahedron model using GNF itself and its embedded version with different $\sigma_{r}$ values. The line charts of the obtained MAE and $E_v$ values are shown in Fig.~\ref{fig:gnf-stability-trend}. These results demonstrate that our approach can significantly boost the stability of the existing mesh denoising methods. 

\subsection{Refinement Step}
\label{subsec:refinement-compare}
\begin{figure} [htbp]
\centering
			\includegraphics[width=\columnwidth]{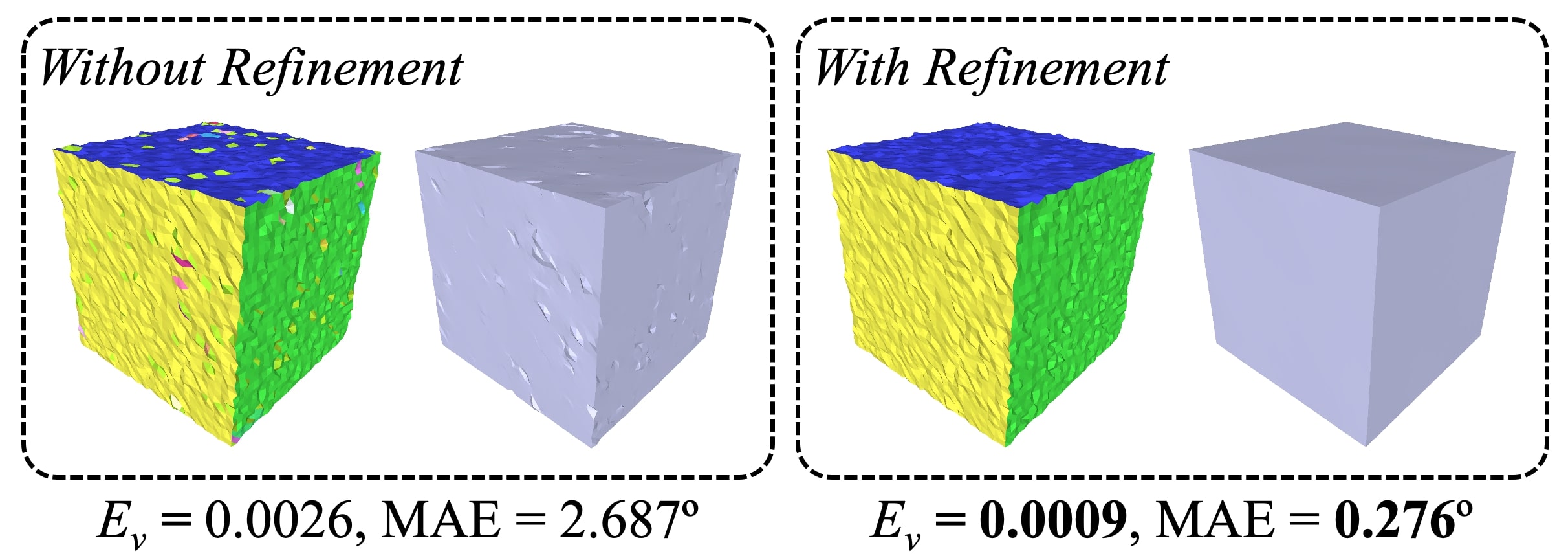}
\caption{Denoising results without and with the cluster refinement step, where the MAE and $E_v$ values are smaller with the refinement step. }
\label{fig:refine}
\end{figure}

As previously mentioned, we apply a region refinement step after the initial segmentation step. This indeed assists in improving the quality of the denoised mesh. As shown in Fig.~\ref{fig:refine}, small clusters are considered as anisotropic surfaces without the refinement step, even if they originally belong to the same region as their surrounding faces. As a consequence, it leaves pits and rough edges on the denoised mesh. In contrast, after applying the refinement step, the denoised shape has cleaner surfaces and tidier edges, achieving lower MAE and $E_v$ values.

\subsection{Limitations}
\begin{figure} [htbp]
\centering
			\includegraphics[width=0.7\columnwidth]{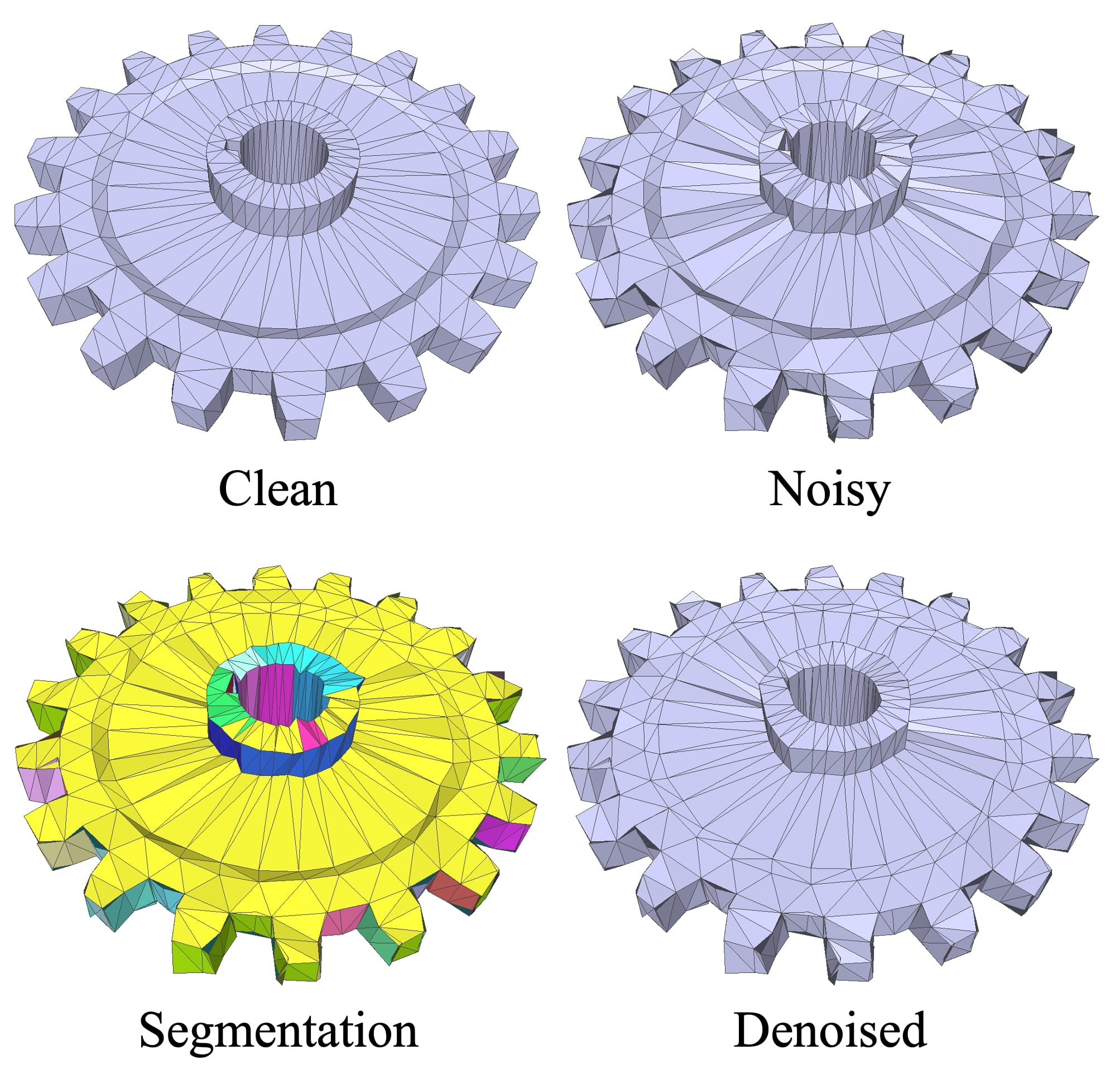}
\caption{A failure case of denoising a Gear model with extremely sparse triangulation, with \cite{Lu2017-L1median} as our denoising backbone.}
\label{fig:limitation-sparse}
\end{figure}

Our method has a few limitations. For example (and similar to other methods), for any model with extremely sparse and irregular triangulation, our segmentation backbone can hardly guide the denoising process. Fig.~\ref{fig:limitation-sparse} demonstrates segmentation and denoising results on a Gear model with extremely sparse and irregular triangulation. Although our edge-based segmentation result can reasonably partition the mesh's geometric regions, the denoised result is still not satisfactory, as the denoised Gear model does not fully reveal its original shape. This can be seen at the edges, where the triangles are folded. In addition, as our method is geometric-based, it is unaware of the semantic segmentation. As for our future works, we would like to design more robust segmentation techniques for noisy 3D shapes. 

\section{Conclusion}
We present a novel segmentation-driven mesh denoising framework, which facilitates the benefits brought by geometric mesh segmentation, and helps with achieving feature-preserving mesh denoising. Moreover, it can be easily integrated with existing mesh denoising methods with demonstrated robustness. Both visual and quantitative results confirm that our method enables better feature-preserving mesh denoising outcomes, especially for models corrupted with severe noise.

{\small
\bibliographystyle{cvm}
\bibliography{cvmbib}
}

\end{document}